\documentclass{aa}
\usepackage{txfonts}
\usepackage{graphicx}
\usepackage[normalem]{ulem}
\begin{document}
%
%
\newcommand{\cm}{\,{\rm cm}}
\newcommand{\cmcube}{\,{\rm cm^{-3}}}
\newcommand{\dyn}{\,{\rm dyn}}
\newcommand{\erg}{\,{\rm erg}}
\newcommand{\Jy}{\,{\rm Jy}}
\newcommand{\Jyb}{\,{\rm Jy/beam}}
\newcommand{\kms}{\,{\rm km\,s^{-1}}}
\newcommand{\mJy}{\,{\rm mJy}}
\newcommand{\mJyb}{\,{\rm mJy/beam}}
\newcommand{\K}{\,{\rm K}}
\newcommand{\kpc}{\,{\rm kpc}}
\newcommand{\Mpc}{\,{\rm Mpc}}
\newcommand{\mG}{\,{\rm mG}}
\newcommand{\mkG}{\,\mu{\rm G}}
\newcommand{\MHz}{\, {\rm MHz}}
\newcommand{\Msol}{\,{\rm M_\odot}}
\newcommand{\p}{\,{\rm pc}}
\newcommand{\radm}{\,{\rm rad\,m^{-2}}}
\newcommand{\s}{\,{\rm s}}
\newcommand{\yr}{\,{\rm yr}}
\newcommand{\Gyr}{\,{\rm Gyr}}

\newcommand{\cs}{c_{\rm s}}
\newcommand{\rc}{r_{\rm c}}
\newcommand{\ncr}{n_\mathrm{cr}}
\newcommand{\nel}{n_\mathrm{e}}
\newcommand{\RM}{\mathrm{RM}}
\newcommand{\PA}{\mathrm{PA}}
\newcommand{\PI}{P}
\newcommand{\RA}{\mathrm{RA}}
\newcommand{\Dec}{\mathrm{Dec}}
\newcommand\sfrac[2]{{\textstyle{\frac{#1}{#2}}}}

\title{Magnetic fields in barred galaxies}
\subtitle{V. Modelling NGC~1365}

\author{D.\ Moss\inst{1}
\and
A.\ P.\ Snodin\inst{2}
\and
P.\ Englmaier\inst{3}
\and
A.\ Shukurov\inst{2}
\and
R.\ Beck\inst{4}
\and
D.\ D.\ Sokoloff\,\inst{5} }

\institute{School of Mathematics, University of Manchester,
Oxford Road, Manchester, M13 9PL, UK
\and
School of Mathematics and Statistics, University of Newcastle,
Newcastle upon Tyne, NE1 7RU, UK
\and
Institute of Theoretical Physics, University of Z\"urich, Winterthurerstrasse 190, 8057 Z\"urich, Switzerland
\and
Max-Planck-Institut f\"ur Radioastronomie,
Auf dem H\"ugel 69, 53121 Bonn, Germany
\and
Department of Physics, Moscow University, 119992 Moscow, Russia
}

\date{Received ...; accepted ...}

\offprints{D.\ Moss }

\titlerunning{Modelling NGC 1365}
\authorrunning{Moss et al.}

\abstract{}   
{            
We present a model of the global magnetic field in the barred galaxy
NGC~1365 based jointly on the large-scale velocity field of interstellar
gas fitted to \ion{H}{i} and \ion{CO}{} observations of this galaxy
and on mean-field dynamo theory. The aim of the paper is to
present a detailed quantitative comparison of a galactic dynamo model with
independent radio observations.}
{            
We consider several gas dynamical models, based on two rotation curves. 
We test a range of nonlinear dynamo models that include plausible variations
of those parameters that are poorly known from observations. Models for the
cosmic ray distribution in the galaxy are introduced in order 
to produce synthetic radio polarization maps that allow direct comparison with
those observed at $\lambda\lambda3.5$ and $6.2\cm$.}
{            
We show that the dynamo model is robust in that the most important
magnetic features are controlled by the relatively well established properties
of the density distribution and gas velocity field. 
The optimal agreement between the synthetic polarization maps and observations
is obtained when a uniform cosmic ray distribution is adopted.
These maps are sensitive to the number density of thermal ionized gas
because of Faraday depolarization effects.
Our results are compatible with the observed polarized radio intensity
and Faraday rotation measure if the degree of ionization is between 0.01 and
0.2 (with respect to the total gas density, rather than to the diffuse
gas alone). We find some indirect evidence for enhanced turbulence in the
regions of strong velocity shear (spiral arms and large-scale shocks in the
bar) and within 1--2\,kpc of the galactic centre. We confirm that magnetic
stresses can drive an inflow of gas into the inner 1\,kpc of the galaxy at a
rate of a few $M_\odot\yr^{-1}$.}
{           
The dynamo models are successful to some extent in modelling the
large scale regular magnetic field in this galaxy.
Our results demonstrate that dynamo models and synthetic polarization maps
can provide information about both the gas dynamical
models and conditions in the interstellar medium. In particular, it seems
that large-scale deviations from energy equipartition (or pressure balance) between
large-scale magnetic fields and cosmic rays are unavoidable. 
We demonstrate that the dynamical effects of magnetic fields cannot be
everywhere ignored in galaxy modelling.     
}


\maketitle

\section{Introduction}
NGC~1365 is one of the best studied barred galaxies. It has been observed
in a broad range of wavelengths, including \ion{H}{i} (Ondrechen \& van der Hulst
\cite{OH89}), molecular gas (Curran et al.\ \cite{C01}), H$\alpha$
(Lindblad \cite{L99}), and the radio
range (Sanqvist et al.\ \cite{SJ95}; Beck et al.\ \cite{Betal05}), in addition to
numerous optical and infrared
observations (see Lindblad \cite{L99} and references therein). Detailed
gas dynamical modelling by Lindblad et al.\ (\cite{LLA96}) provided quantitative
models for the gravity and gas velocity fields in this galaxy that fit the \ion{H}{i}
and, to some extent, the CO observations.

The aim of this paper is to add to these efforts by the inclusion of
magnetic fields. The gas dynamical model of the galaxy then can be
tested against independent radio data, which were not included in the
construction of the model. Of course, this involves an additional piece of
theory and some further assumptions (concerning, e.g., the applicability of dynamo
theory to galaxies
and uncertainties in some dynamo parameters). Some features of the
dynamo theory certainly are not understood well enough.
However, we demonstrate that
gross features of the model galactic magnetic field -- at least in barred galaxies
where the shear in the large-scale velocity is the dominant
induction effect -- are rather
insensitive to the poorly known details of the dynamo system (most
importantly, the $\alpha$-coefficient). Therefore, we can plausibly
constrain the freedom within the dynamo models, and so draw conclusions about
the interstellar medium in barred galaxies.

We find fair agreement between radio polarization
observations and the magnetic field obtained as a solution of the mean-field
dynamo equations, using velocity and density fields obtained from gas dynamical
simulations, although the distribution of polarized intensity is reproduced
better than that of polarization angles. Our models also support the idea that 
interstellar turbulence is enhanced in the vicinity of dust lanes near the bar major axis,
and that the
energy density of cosmic rays can depend only weakly on position in the
galaxy, thus deviating significantly from equipartition with interstellar
magnetic field. As a result, radio polarization observation and modelling of
magnetic fields are important ingredients of both the theory and observations
of barred galaxies.
This work resembles quite strongly an earlier study of another barred galaxy,
NGC~1097 (Moss et al. \cite{MS01}), but represents a significant improvement
in that we now use a dynamical model that specifically models NGC~1365, rather than
the generic dynamical model adopted for NGC 1097.
Also, the dynamo model we use here is fully three dimensional,
whereas that of Moss et al.\ (\cite{MS01}) used the `no-$z$' approximation
to remove explicit dependence on the vertical coordinate.
Broadly comparable studies have also been published by Otmianowska-Mazur et al.
(\cite{O-M02}), Soida et al. (\cite{S06}) and Vollmer et al. (\cite{V06}).

\begin{figure*}
\begin{center}
\includegraphics[width=0.49\textwidth]{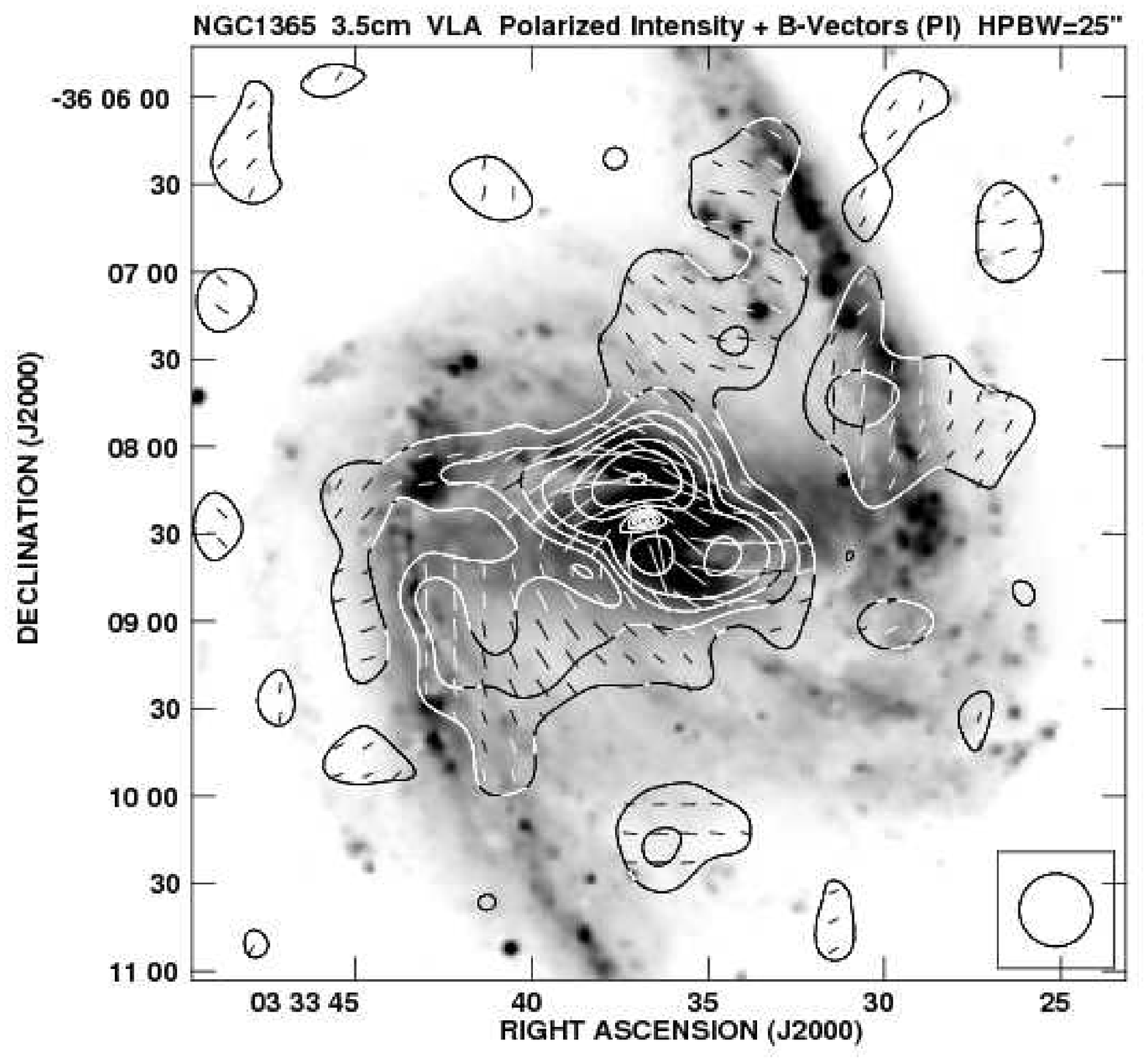}
\hfill
\includegraphics[width=0.49\textwidth]{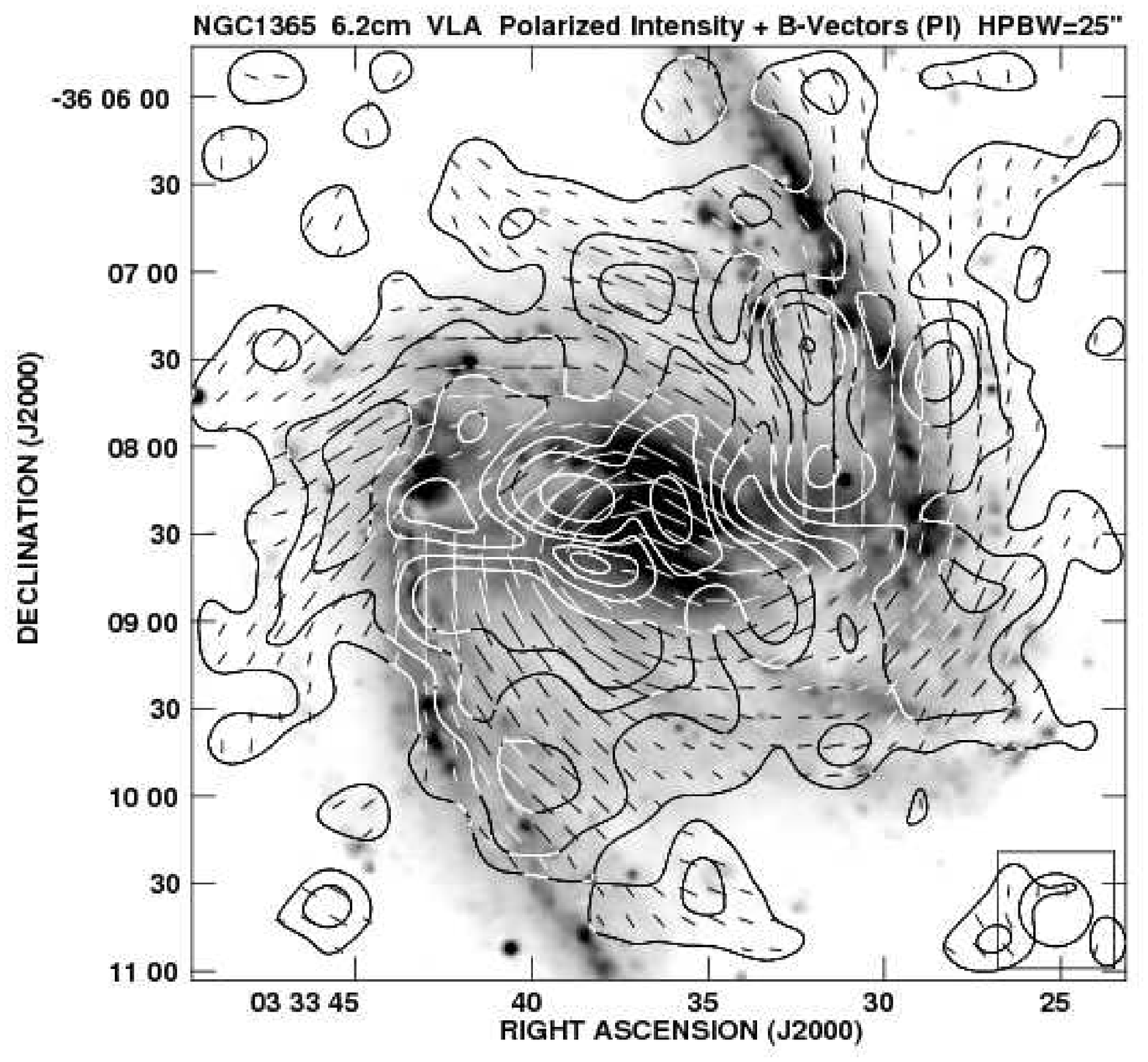}
\end{center}
\caption{\label{PIobs} The polarized intensity contours and magnetic vectors of
the polarized radio emission at the wavelengths $\lambda3.5\cm$ (left hand panel)
and $\lambda6.2\cm$ (right hand panel) (both smoothed to a resolution $25\arcsec$;
the beam size is shown in the lower right of each panel), superimposed onto an
ESO optical image of NGC 1365, kindly provided by P.~O.~Lindblad.
The contour levels are $1, 2, 3, 4, 6, 8, 12,
\ldots$ times $30\,\mu\mathrm{Jy/beam}$ at $\lambda3.5\cm$ and
$40\,\mu\mathrm{Jy/beam}$ at $\lambda6.2\cm$;
the r.m.s. noise is
$15\,\mu\mathrm{Jy/beam}$ at $\lambda3.5\cm$ and $14\,\mu\mathrm{Jy/beam}$ at
$\lambda6.2\cm$.}
\end{figure*}

\section{The observed magnetic structure}
\label{observations}
NGC~1365 was observed in total and polarized radio continuum with the VLA DnC
array at $\lambda$3.5~cm and $\lambda$6.2~cm. The full details and the maps at
15\arcsec\ and 25\arcsec\ angular resolution are given in Beck et al.\
(\cite{Betal05}). The total radio
intensity (a measure of total magnetic field strength and thermal emission) follows
well the optical bar and the spiral arms. According to the observed spectral indices,
the thermal fraction is about 20\% at $\lambda$ 6.2~cm.

The polarized emission (Fig. \ref{PIobs}) is strongest in the central region and
inner bar, but decreases rapidly towards the outer bar. There is also significant
polarized emission between the bar and the spiral arms. No concentration in
the spiral arms can be detected. At $\lambda6.2\cm$, where the sensitivity is
highest, the polarized emission forms a smooth halo around the bar. The degree of
polarization is low in the bar and spiral arms, indicating that the turbulent
magnetic field dominates in the regions of high gas density and strong star
formation, while the regular field is strong between the bar and the spiral arms.
At $\lambda3.5\cm$, most of the extended polarized emission outside the bar is
lost in the noise because of the steep synchrotron spectrum. Furthermore, the
sensitivity of the VLA to extended structures is reduced for scales beyond 3
arcminutes at $\lambda3.5\cm$, which affects the visibility of the large-scale
polarized emission in NGC~1365, while at $\lambda6.2\cm$ the critical limit is
5 arcminutes and so does not affect our observations.

The peak polarized intensity is 368 mJy per beam at $\lambda3.5\cm$ in the
massive dust lane northeast of the centre (see Beck et al.\ \cite{Betal05}). The
fractional polarization is 0.8. At the same position the $\lambda6.2\cm$ map
reveals a local {\it minimum\/} with  polarized intensity of 150 mJy/beam,
corresponding to a fractional polarization of only 0.2, which is near the
expected contribution from instrumental polarization by the bright nuclear
region. This indicates that strong depolarization occurs at $\lambda6.2\cm$ in
the central region, by a factor of at least 4. In the bar and spiral arms the
depolarization factor is 2--3 (Beck et al.\ \cite{Betal05}).

Polarized emission can emerge from coherent, regular magnetic fields or from
anisotropic random
magnetic fields; these possibilities can be distinguished with the help of Faraday
rotation
measures. In NGC~1097,
anisotropic fields dominate in the bar region (Beck et al.\ \cite{Betal05}). However,
due to the weak polarized intensity in NGC~1365, the observations available
cannot provide a large-scale map of Faraday
rotation, so that the relative
contributions of coherent and anisotropic random magnetic fields remains unclear.
\begin{figure*}
\begin{center}
\hfill
\includegraphics[width=0.32\textwidth]{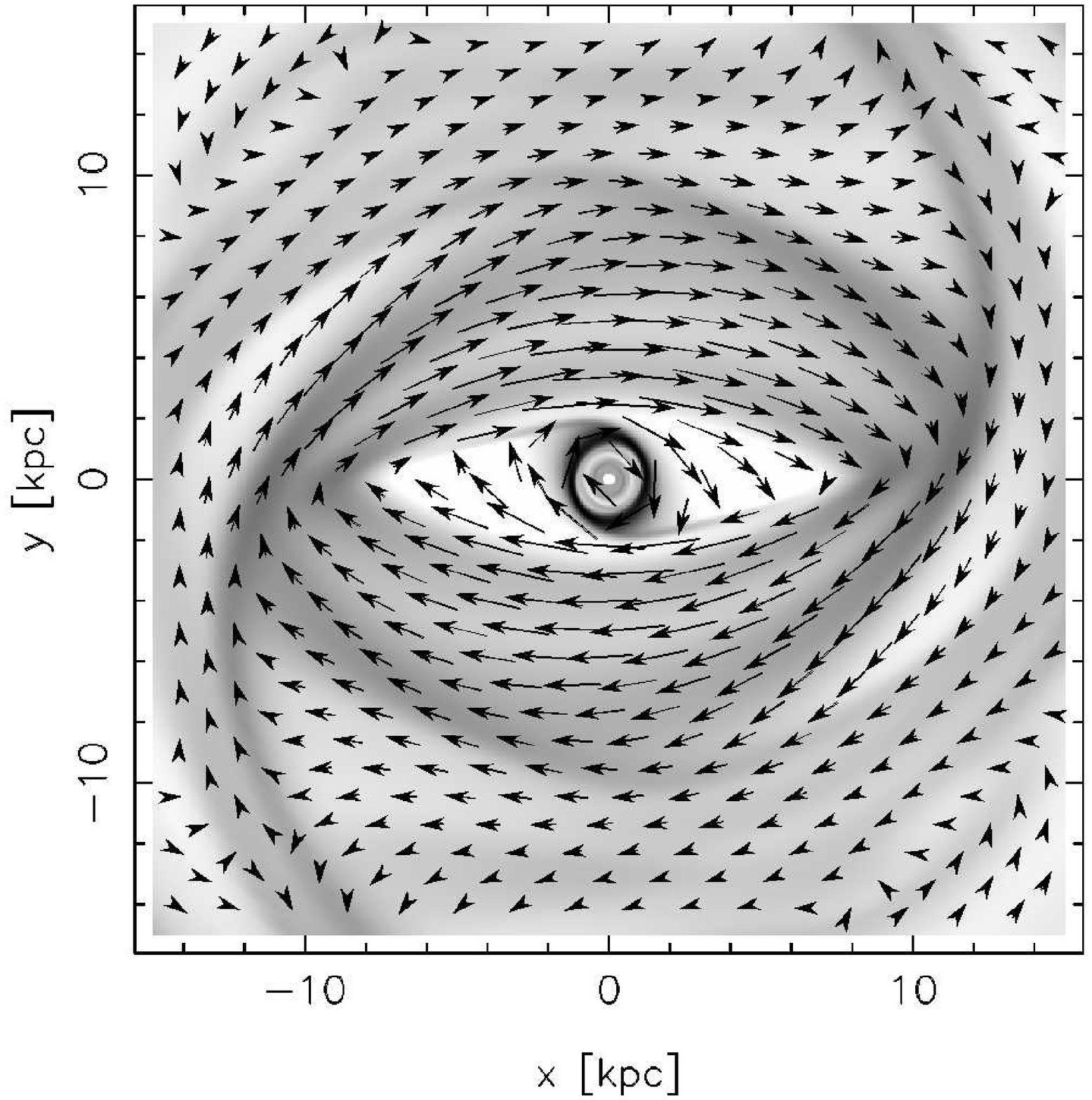}
\hfill
\includegraphics[width=0.32\textwidth]{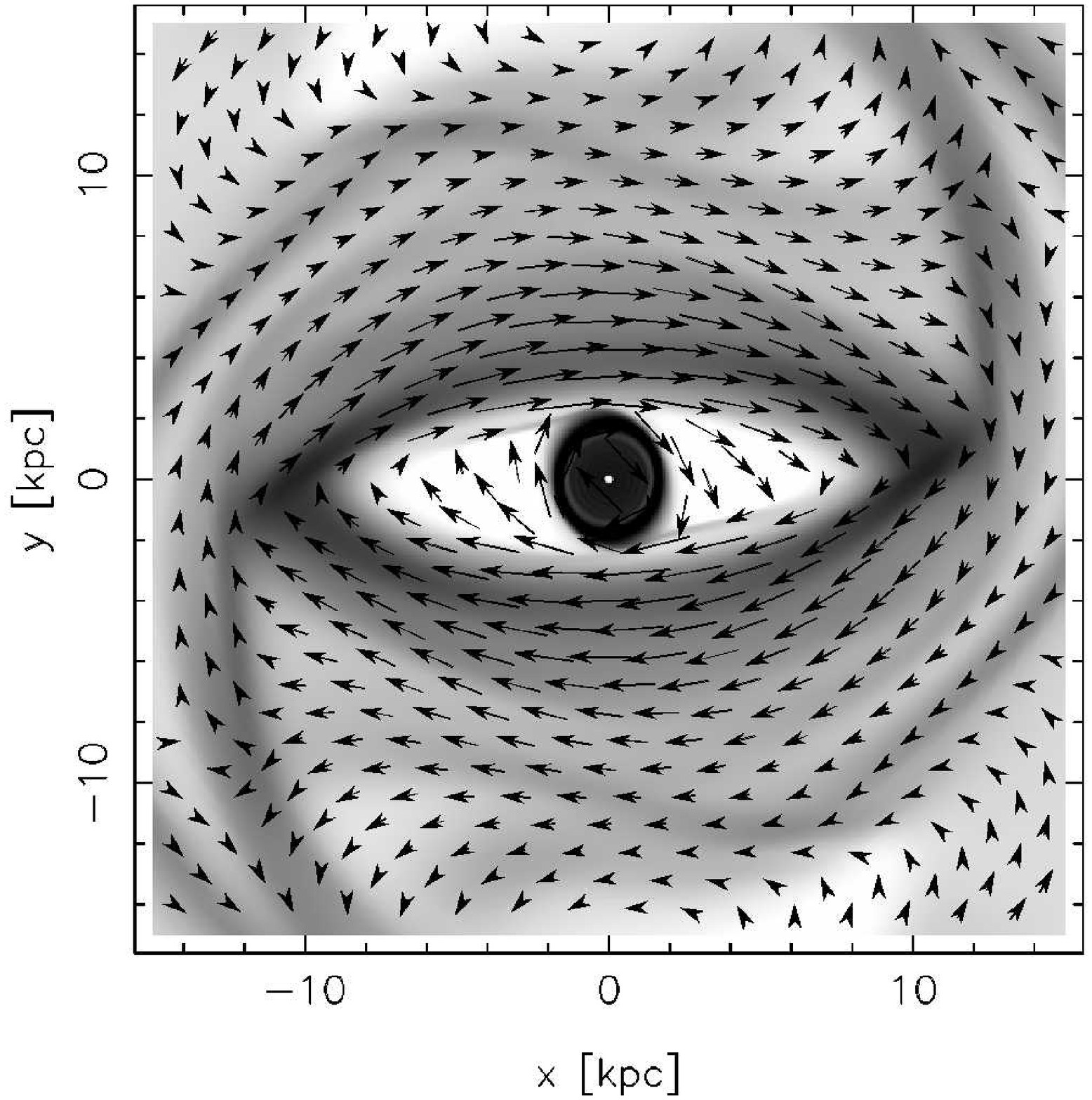}
\hfill
\includegraphics[width=0.32\textwidth]{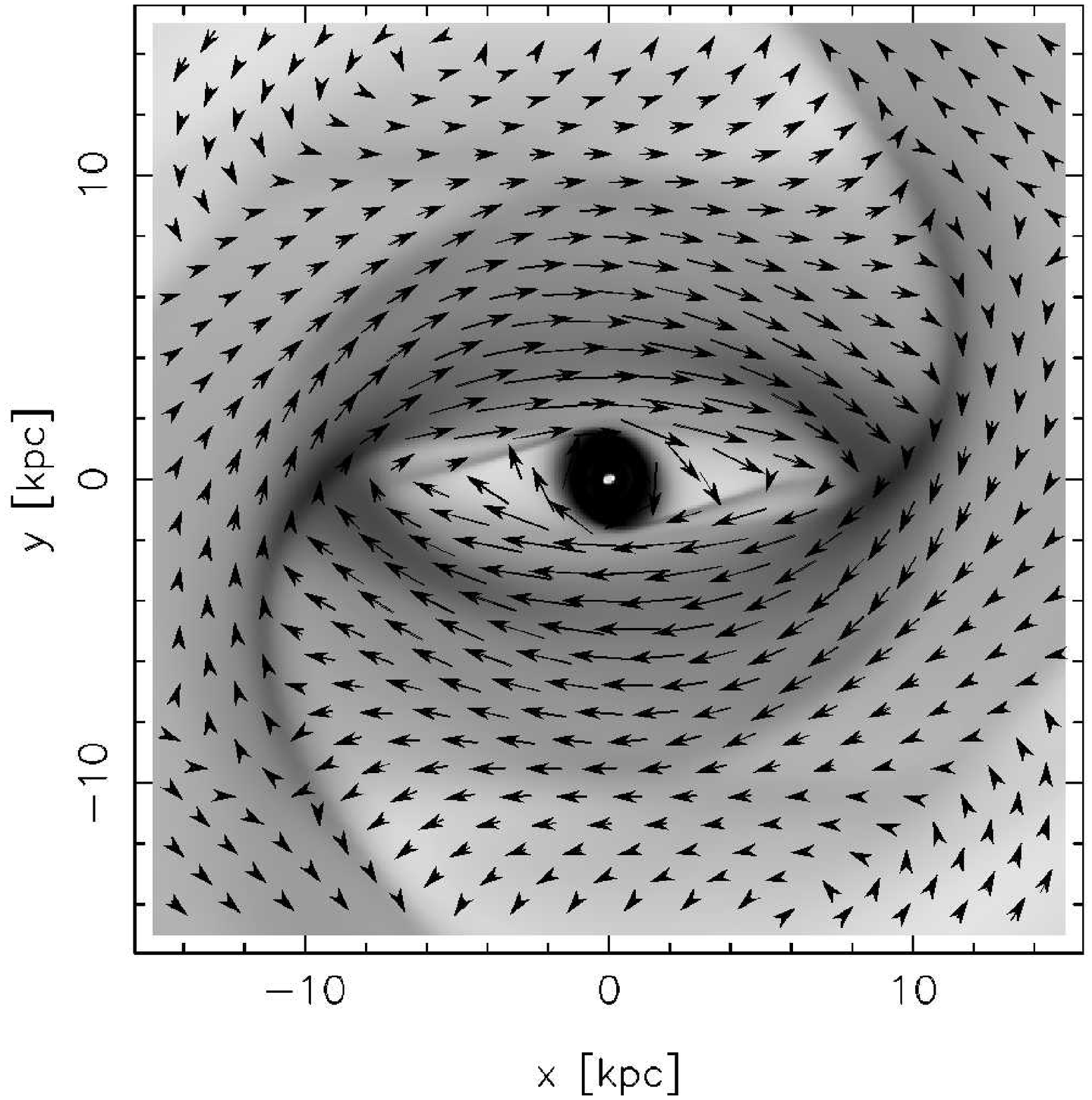}
\hfill
\end{center}
\caption{
The model gas density with superimposed velocity
vectors in the reference frame corotating with the bar, in gas dynamical models
based on {\bf(a)} the rotation curve of the LLA model
with  $\cs=10\kms$
(left hand panel), and the rotation curve of Sofue et al.\
(\cite{S99}) with
{\bf(b)} $\cs=10\kms$
(middle panel) and {\bf(c)}  $\cs=30\kms$
(right hand panel), with $\cs$ the sound speed.
Shades of grey represent the logarithm of gas density (darker shades corresponding
to larger values), with each shade corresponding to the same density in each
panel. Note the smaller density
contrast in the bar region in the model with higher speed of sound (panel c).
}
\label{gas_1030}
\end{figure*}

\section{The model}
\subsection{Gas dynamical models of NGC~1365} \label{gasdyn} We reproduced the
gas dynamical model of Lindblad, Lindblad \& Athanassoula (\cite{LLA96}) using
their gravitational potential `BSM' kindly provided to us by P.~O.~Lindblad.
This potential (the `LLA model' in the following) includes the gravitational
potentials of the disc and spiral arms and was derived from the nonaxisymmetric
part of the deprojected $J$-band image. Their best fit parameters are
$A_\mathrm{bar}=1.2$ and $A_\mathrm{spiral}=0.3$ for the relative contributions
of the bar and spiral arms.
The model rotation curve fits the \ion{H}{i} rotation curve for galactocentric
distances $r>120\arcsec$ and gives reasonable resonance locations inside this
radius.
Various versions of the LLA model used the bar angular velocity of
$\Omega_\mathrm{p}= 18\kms\kpc^{-1}$ (model BSM) and $17\kms\kpc^{-1}$ (model
BSM2), with the corotation radius close to $14\kpc$ in both cases.

The full gravitational potential of the LLA model is obtained from two independent
observations: (i) the \ion{H}{i} rotation curve, used to fix the total radial
mass distribution of the galaxy including dark matter, and (ii) the $J$-band
data, tracing the stellar mass distribution, which is only used to derive
(after deprojection) the amplitude of nonaxisymmetric perturbations in the
disc plane. The latter cannot be used to derive the rotation curve reliably
because of the presence of dark matter, and the former also can be misleading
when the gas flow is significantly nonaxisymmetric.

Lindblad et al.\ (\cite{LLA96}) adopted $20\Mpc$ ($1\arcsec = 97\p$) for
the distance of NGC~1365, but we adjusted the model to a distance
of 18.6\,Mpc ($1\arcsec=90\p$) (Lindblad \cite{L99}).

\begin{figure}
\begin{center}
\includegraphics[width=0.47\textwidth]{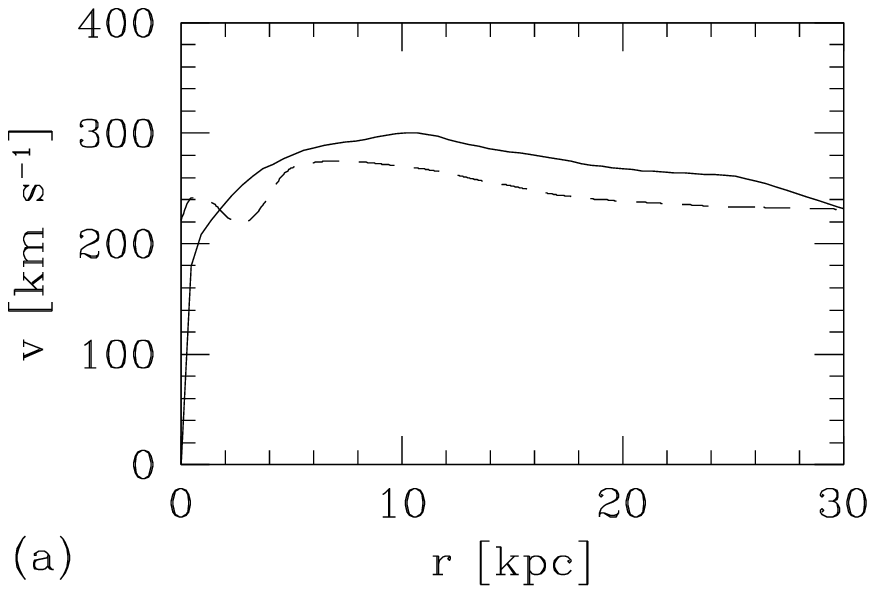}\\
\includegraphics[width=0.47\textwidth]{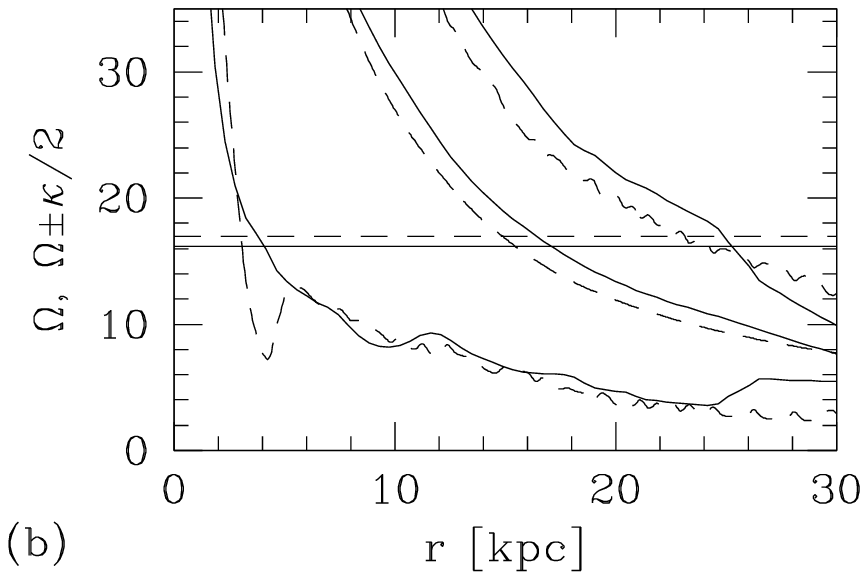}
\end{center}
\caption{{\bf(a)}: The rotation curves used in the paper: that from
Lindblad et al.\ (\cite{LLA96}) (solid;
as in Fig.~\ref{gas_1030}a),
and one more consistent with
more recent CO observations (Sofue et al.\ \cite{S99}) (dashed;
as in Fig.~\ref{gas_1030}b,c).
The plot assumes the distance of NGC~1365 to be 20\,Mpc as in Lindblad et al.\
(\cite{LLA96}).
The radius of corotation is $R_\mathrm{c}\approx14\kpc$.
{\bf(b)}: The linear resonance diagram for the rotation curves shown in
(a) with the same line style. From bottom to top: $\Omega-\kappa/2$, $\Omega$, and
$\Omega+\kappa/2$ in units of $\kms\kpc^{-1}$.
The resonances are located at
the intersections with the horizontal lines corresponding
to $\Omega_\mathrm{p}=16.16\kms\kpc^{-1}$ (solid) and $17\kms\kpc^{-1}$ (dashed).
The small scale structure in the $\Omega\pm\kappa/2$
curves is an artefact of plotting.
}
\label{rotcurves}
\end{figure}

Isothermal gas dynamical models were calculated using
the code ZEUS 2D, published by Stone \& Norman (\cite{SN92}),
and we found a close match to the model of
Lindblad et al.\ (\cite{LLA96}). However we did not attempt to take into account
the warp in the outer disc, as we are mostly interested in the inner region.
Our basic models, illustrated in Fig.~\ref{gas_1030}b,c have the bar
angular velocity $\Omega_\mathrm{p}=16.16\kms\kpc^{-1}$ and the corotation
radius at $R_\mathrm{c}=15.5\kpc$; we also considered a model
(Fig.~\ref{gas_1030}a) with  $\Omega_\mathrm{p}=17\kms\kpc^{-1}$ and
$R_\mathrm{c}=16.3\kpc$. The angular velocity of the spiral pattern is taken
to be equal to that of the bar.
For reasons explained below in Section~\ref{results}, the resulting gas density in the
bar region was too low to reproduce the observed magnetic field
within the dynamo model. The gas density in the LLA model can be argued to be
underestimated inside the corotation radius because the rotation curve used
had poor resolution, and underestimates the depth of the
potential well. We derived our
basic model from the LLA model, by
replacing the rotation curve used by Lindblad et al.\ (\cite{LLA96}) with the
more recent CO rotation curve of Sofue et al.\ (\cite{S99}). This modified
model was much better able to reproduce the observed magnetic field, while
remaining in agreement with
the overall morphology of the molecular gas distribution.
A significant difference is that there is more material in the central
regions when Sofue's rotation curve is used.
The rotation curve
used here is shown in Fig.~\ref{rotcurves}a,
with the positions of resonances illustrated in Fig.~\ref{rotcurves}b.

We also studied the dependence of the gas dynamics and magnetic field on the
sound speed adopted in the isothermal gas model. This parameter is
uncertain in our models for several reasons. Englmaier \& Gerhard
(\cite{EG97}) showed that the large-scale gas distribution in isothermal gas
flow models of barred galaxies can depend on the sound speed, even if the
pressure forces are negligible.
Since the position, and even existence, of
shocks depends on Mach number, the global gas flow configuration can change as
a result of a relatively small change in the speed of sound.
Different parts of the
multi-phase interstellar medium (ISM) may not follow the same global gas flow.
Different numerical methods have been shown to represent different aspects of
the ISM with varying success. Sticky particle methods, for example, model
better the clumpy ISM, while grid-based methods give a better description of
the shocks and the smooth gas component.

The global magnetic field depends on the gas flow via Eqs~(\ref{mfd}) and
(\ref{alpha});
however, it is not {\em a priori\/} clear which component of the ISM carries
the magnetic field and, therefore, what is the appropriate sound
speed of the gas. We have considered models with the speed of sound equal to
10 and $30\kms$ (see Sect.~\ref{TEOTSS}).

Our magnetic field model also relies on the gas density obtained from gas
dynamical simulations together with the velocity field; this is discussed in
Sect.~\ref{TDM} -- see Eq.~(\ref{dens}).

\begin{table*}
\caption{\label{run_params}Parameters of models discussed in the text, as
defined in Sect.~\ref{code}. In all the models, the angular speed of the
bar is $\Omega_\mathrm{p}=16.16\kms\kpc^{-1}$ with the corotation radius at
$15.5\kpc$.}
\begin{center}
\begin{tabular}{ccccccc}
\hline\hline
Model &$R_\alpha$ &$\eta_0$ & $q_\eta$ &$r_\eta$ &$f_\eta$ &$\cs$ \\
      & &$[10^{26}\cm^2\s^{-1}]$  &    &[kpc]    &         &$[\!\kms]$ \\
\hline
1  & 3.0 & 1.0 & 3 & 3.0 & 0 &10 \\ 
2  & 3.0 & 1.0 & 3 & 1.5 & 2 &10 \\
3  & 0.0 & 1.0 & 3 & 1.5 & 2 &10 \\
4  & 2.7 & 2.5 & 3 & 1.5 & 2 &10 \\
5  & 3.0 & 2.0 & 3 & 1.5 & 2 &10 \\
6  & 3.0 & 1.0 & 3 & 1.5 & 2 &30 \\
\hline
\end{tabular}
\end{center}
\end{table*}

\subsection{The dynamo model}\label{TDM}
Dynamo models, specifically simple mean-field turbulent dynamos, are remarkably
successful in explaining the observed features of galactic magnetic fields
(see, e.g., Ruzmaikin et al.\ \cite{RSS88}; Beck et al.\ \cite{B96}; Widrow
\cite{W02} for reviews). Despite the fact that the
nonlinear behaviour of turbulent dynamos is still controversial, mean-field
models provide a remarkably reliable empirical description of large-scale
(regular) galactic magnetic fields (Shukurov \cite{S04}).
Fortunately, dynamo
solutions for galaxies are quite insensitive
to those parameters that are poorly known, such as the form of the
$\alpha$-effect and even, to a lesser extent, the turbulent magnetic diffusivity.
This is especially true of models for barred galaxies where large-scale
velocity shear plays a dominant role in determining magnetic field structure
(Moss et al.\ \cite{MK98,MS01}); then the primary role of the
$\alpha$-effect is to maintain the field against decay.

\label{code}
Our model can be regarded as a development of the dynamo model of Moss
et al.\ (\cite{MS01}),
used to model the large-scale magnetic field in a generic barred galaxy.
We now introduce further elaborations required to reproduce the basic
features of the global magnetic pattern in NGC~1365.
We solve the standard mean field dynamo equation for the large-scale, regular
magnetic field $\vec B$
\begin{equation}
\frac{{\partial\vec B}}{\partial t}= \nabla\times\left({\vec u}\times{\vec B}
+\alpha{\vec B}-\sfrac{1}{2}\nabla\eta\times{\vec B}
-\eta\nabla\times{\vec B}\right), \label{mfd}
\end{equation}
in three spatial dimensions, using Cartesian coordinates $(x,y,z)$, where $x$ and
$y$ are horizontal dimensions, and the disc midplane is at $z=0$.
Here $\alpha$ and $\eta$ are the turbulent transport coefficients
responsible for the $\alpha$-effect and turbulent magnetic diffusion,
respectively, $\vec{u}$ is the large-scale velocity field,
and the term with $\nabla\eta$ allows for the turbulent diamagnetism
associated with the spatial variation of the turbulent diffusivity
(Roberts \& Soward \cite{RS75}).
In our standard case, our computational domain covers the region
$-L\le (x,y)\le L$, $0\le z\le aL=z_{\max}$, where $a$ is the domain's aspect
ratio. We take a mesh of size $n_{x} \times n_{y} \times n_{z}$, with uniform
spacing in the horizontal directions and also, separately, vertically. The
maximum resolution readily available to us was $n_{x}=n_{y}=200$, $n_z=31$,
and in order to resolve satisfactorily the solutions we took $L=15\kpc$ and
$a=0.12$, so $z_{\rm max}=1.8\kpc$. (Thus we study only the inner part of
this unusually large barred galaxy.)
The total thickness of gas layer that hosts the large-scale magnetic field
is taken as $2h=0.9\kpc$, compatible with the thickness of the diffuse
warm gas in the Milky Way.
Our procedure is to time-step the $x$ and $y$ components of Eq.~(\ref{mfd}),
and then to use the condition $\nabla\cdot{\vec B} = 0$ to update $B_z$. We
restrict ourselves to solutions of even (quadrupolar) parity with respect to
the disc plane $z=0$, and so the latter step is straightforward, given that
$B_z=0$ at $z=0$. This is the same procedure used in the three-dimensional
galactic dynamo models described in Moss (\cite{M97}), except that cylindrical
polar coordinates were used there.

In Eq.~(\ref{mfd}), $\alpha$ parameterizes the dynamo action of the
interstellar turbulence, and $\eta$ is the turbulent magnetic diffusivity. We
assume both of these quantities to be scalars (rather than tensors)
and, in order to obtain a steady
state with saturated dynamo action, introduce a simple $\alpha$-quenching
nonlinearity into the problem, writing
\begin{equation}\label{alpha}
\alpha = \frac{\alpha_0}{1+\xi{\vec B}^2/B_\mathrm{eq}^2}\;,
\qquad
B_\mathrm{eq}^2=4\pi\rho({\vec r})v_\mathrm{t}^2\;,
\end{equation}
\begin{equation}
\alpha_0=\alpha_*\frac{\Omega(r)}{\Omega_0}f(z)\;,
\label{alpha0}
\end{equation}
with
\[
f(z)= \left\{ \begin{array}{ll} \displaystyle \sin(\pi z/h)\;,
&|z|\le h/2\;,\\[7pt]
\displaystyle
\left[\cosh\left(2{|z|}/{h}-1\right)^{2}\right]^{-1}\mathrm{sgn}\,{z}\;,
&|z|>h/2\;.
\end{array} \right.
\]
Here $\Omega_0$ is a typical value of $\Omega$, $B_\mathrm{eq}$ is the
magnetic field strength corresponding to equipartition between magnetic and
turbulent kinetic energies, and $\alpha_*$ is a constant, which we can adjust.
Quite arbitrarily, we adopt $\Omega_0=\Omega$ at $r=3\kpc$, and
Eq.~(\ref{alpha0}) shows that
$\alpha_*$ is the maximum value of $\alpha$ at this radius.
Thus we are
assuming that the large-scale magnetic field significantly reduces the
$\alpha$-effect when its energy density approaches that of the turbulence; the
constant $\xi$ is introduced to suggest formally some of the uncertainty about
the details of this feedback. The dependence of $\alpha$ on height, defined by
$f(z)$, is implicitly odd with respect to the midplane,
with $|\alpha|$ increasing with $|z|$ from 0 at $z=0$ to a
maximum at $|z|=h/2$, and then decreasing to zero as $|z|\to\infty$ (remembering
that we only explicitly model the region $z\ge 0$).
Because of the symmetry of Eqs~(\ref{mfd}) and (\ref{alpha}), if $\vec{B}$
is a solution, then $-\vec{B}$ is also a solution.

We take $\xi=O(1)$, assuming that there is no catastrophic
$\alpha$-quenching (Brandenburg \& Subramanian \cite{BS05}). The
models were computed with $\xi=1$, and the field strength
then scales as $\xi^{-1/2}$. The turbulent
speed that enters $B_\mathrm{eq}$ is taken to be equal to the speed of sound
as adopted in the gas dynamical model. The gas density $\rho(x,y,0)$ is taken from
the gas dynamical model described in Sect.~\ref{gasdyn}. We extend this away
from $z=0$ by writing
\begin{equation}
\rho(x,y,z)=\frac{\rho(x,y,0)}{\cosh({|z|}/{h})}\;. \label{dens}
\end{equation}
The magnitude of the gas density is relatively unimportant in our model (where
the Lorentz force is not included into the Navier--Stokes equation) as
it affects only the magnitude of the magnetic field in the steady state, via
Eq.~(\ref{alpha}), but not its spatial distribution. The only aspect where the
magnitude of gas density plays a role is the Faraday depolarization and,
hence, the modelled distribution of polarized intensity. This effect is,
however, relatively weak at $\lambda=3$--$6\cm$ and it is plausible that other
depolarization effects (e.g., Faraday dispersion) are more important in the
real galaxy. The gas density in our model is shown in Fig.~\ref{B2}.

The gas velocity in the plane $z=0$, ${\vec u}(x,y,0)$, is also taken from
the gas dynamical model. For convenience, we split
this into rotational and non-circular parts,
\begin{equation}
{\vec u}(x,y,0)=\Omega(r)r\widehat{\vec{\phi}} + {\vec v}(x,y,0),
\label{vsplit}
\end{equation}
respectively, where $r=(x^2+y^2)^{1/2}$ is axial distance.

We then introduced two significant modifications. We found that, in the gas
dynamical model, $\Omega(r)$ increases very rapidly towards the rotation axis
(very approximately, as $1/r$). The gas dynamical model model appears
to handle this feature satisfactorily, but it
causes significant numerical problems for the dynamo code
at attainable numerical resolution.
Thus $\Omega$ was softened by introducing an explicit parabolic profile
within a radius of $2.1\kpc$, with the maximum of $\Omega$ truncated to
$110\kms\kpc^{-1}$ (as compared to $1730\kms\kpc^{-1}$ at $r=0.013\kpc$, the smallest distance from the axis in the gas dynamical model used).
This modification can be expected to reduce the magnetic field strength in
regions close to the galactic centre, but as this region is not well resolved
by the radio observations, we cannot in any case make a comparison between
these and the computed magnetic field.

Further, we continued the velocity field above the disc by introducing
$z$-dependence into the horizontal velocity components via
\begin{equation}
{\vec u}(x,y,z)=\frac{{\vec
u}(x,y,0)}{\cosh({|z|}/{1.2\kpc})}\;,
\end{equation}
and $u_z=0$ everywhere.

In order to model a galaxy surrounded by near-vacuum, we allow the magnetic
diffusivity to become large high in the halo (Sokoloff \& Shukurov
\cite{SS90}),
\[
\eta=\eta_0
\left\{
\begin{array}{ll}
\displaystyle
1\;, &|z|\le h\;,\\[7pt]
1+(\eta_1-1)\left[1-\exp\left(-
\displaystyle\frac{|z|-h}{1.5\kpc}\right)\right]^2\;,
&|z|>h\;, \end{array} \right.
\]
where $\eta_0$ and $\eta_1$ are constants; thus
$\eta=\eta_0$ near the disc midplane and  $\eta\rightarrow \eta_0\eta_1$
in the halo region ($|z|>h$). We adopted a nominal $\eta_1=2$ -- larger values
led to numerical difficulties. A conventional
value of $\eta_0$ is $10^{26}\cm^2\s^{-1}$; however, we also considered models
with values larger than that -- see Table~\ref{run_params}. In order to
reproduce polarized radio maps of NGC~1365 in sufficient detail, we had to
introduce further spatial variation in $\eta$. Following Moss et al.\
(\cite{MS01}), we have assumed that the turbulent diffusivity is enhanced
by the shear of the nonaxisymmetric velocity according to
\[
\eta_0\propto\left(1+f_\eta \frac{S}{S_\mathrm{max}}\right)\;,
\qquad
S=\left|\frac{\partial u_x}{\partial y}\right|
+\left|\frac{\partial u_y}{\partial x}\right|\;,
\]
where $S_\mathrm{max}$ is the maximum value of $S$. The effect of
$f_\eta\neq0$ is, firstly, to broaden magnetic structures near the
spiral arms,
and, secondly, to reduce the central peak of magnetic field. The values of
$f_\eta$ adopted are shown in Table~\ref{run_params}. We did not consider a
similar enhancement in $\alpha$ as Moss et al.\ (\cite{MS01}) found it to be
unimportant. The values of $f_\eta$ that were sufficient to produce realistic
magnetic fields in spiral arm were still too small to reduce the central
maximum of magnetic field to an acceptable level. Therefore, we
introduced an additional enhancement of $\eta$ in the central part of the
galaxy, multiplying $\eta_0$ by $q_\eta\exp(-r^2/2r_\eta^2)$; the
values of $q_\eta$ and $r_\eta$ are given in Table~\ref{run_params} for each
model studied.

Clearly, we have made a number of rather arbitrary choices, in particular when
extending the two dimensional gas dynamical model into three dimensions. Our
overall impression from a substantial number of numerical experiments is that
the overall nature of our results does not depend very strongly on these
choices.

At $z=z_{\rm max}$, and on $x,y=\pm L$, the boundary conditions are
$B_x=B_y=0$. On $z=0$, ${\partial B_x}/{\partial z}={\partial B_y}/{\partial
z}=0$, $B_z=0$, and so the integration of $\nabla\cdot{\vec B}=0$ gives the
values of $B_z$ on the other boundaries. These are conservative boundary
conditions on $B_x$ and $B_y$, in that they will increase the field gradients
and thus raise the threshold for dynamo action to occur.

We nondimensionalize the problem in terms of the length
$L=15\kpc$, time $h^2/\eta_0$
and magnetic field $B_\mathrm{eq}$. Given that the velocity field, including
the angular velocity, is given by the dynamical model, the only
free dynamo parameter is $\alpha_*$; the corresponding dimensionless
parameter is
\begin{equation}
R_\alpha=\frac{\alpha_*h}{\eta_0},
\end{equation}
where $\alpha_*$ is defined in Eq.~(\ref{alpha0}).
The dynamo action prevents magnetic field from decay for
values of $R_\alpha$ exceeding about $1$ for $\eta_0=10^{26}\cm^2\s^{-1}$;
the critical value of $R_\alpha$ increases roughly proportionally to $\eta_0$.

Henceforth, we will use dimensionless variables, unless explicitly otherwise
stated; the units of gas number density and magnetic field strength are
$44\cm^{-3}$ and $30\mkG$, respectively.

\section{Results}
\label{results}

\begin{figure}
\begin{center}
\includegraphics[width=0.47\textwidth]{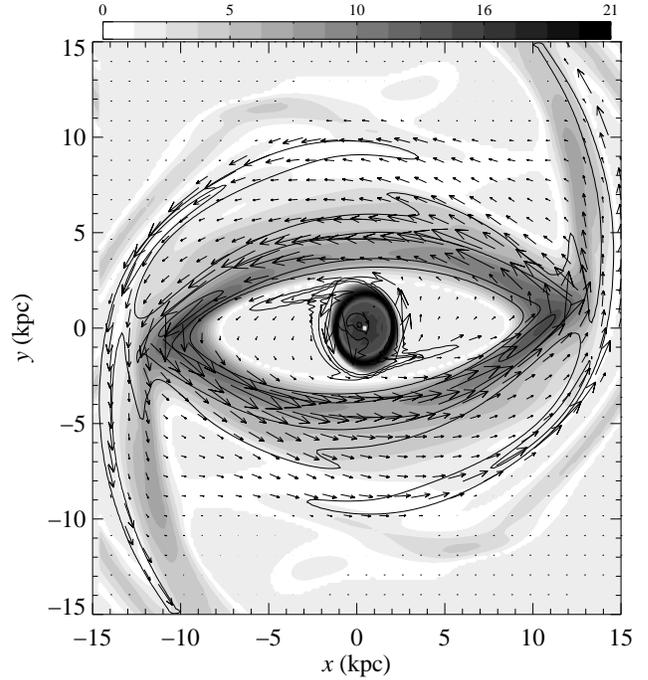}
\end{center}
\caption{Energy density contours and vectors of the regular magnetic field
$\vec{B}$ from Model 2 (see Table~\ref{run_params}),  both
at $z=0$, are shown together with gas number density
represented with shades of grey.
The contours shown correspond to approximately 0.1, 0.6 and 3.0 times the r.m.s.\
value;
the length of the vectors is proportional to $B^2$.
The scale bar at the top of the frame refers to the gas number density
in the units of hydrogen atoms per cm$^3$.
}
\label{B2}
\end{figure}

The gas dynamic and dynamo models described above together yield the gas density and
the distribution of the large-scale magnetic field in the galaxy.  The
distribution of magnetic field in the galaxy plane resulting from Model 2
(introduced in Table~\ref{run_params}), which we argue below to be our best
model, is shown in Fig.~\ref{B2}. Given the distribution of the cosmic rays,
we can now construct synthetic radio observables in order to
assess the quality of the model. We have computed synthetic radio polarization
maps at wavelengths of 3.5\,cm and 6.2\,cm using the dynamo generated magnetic
field and the gas density, and compared them with the observed radio maps.
Details of this procedure are given in Appendix~\ref{synthetic_radio_maps}.
Since we do not model turbulent magnetic fields, we are unable to calculate the
total radio intensity and to estimate the degree of polarization from the
model.

We considered several models for the number density of cosmic rays, $\ncr$,
which we discuss in Sect.~\ref{n_cr_models}. For all but one of the dynamo
models listed in Table~\ref{run_params}
 we find that the simplest possible choice,
$\ncr=\mbox{const},$ provides the best fit to the observed data, regardless of
the other quantities adopted. The larger value of $r_\eta$ in
Model~1 produces a relatively weak magnetic field throughout a large
central region compared with that at the ends of the bar. In order
to fit the observed central peaks of polarized intensity, $\PI$,
an implausible non-uniform distribution of $\ncr$ would be required in this
model. Specifically, the cosmic ray distribution required to reconcile this
model with observations would have a high peak within $3\kpc$ of the centre
where magnetic field strength is minimum. For the other dynamo models, any plausible
non-uniform distribution of
$\ncr$ produces too strong a central maximum of $\PI$ relative to all other
structures. In particular, the polarized intensity in the spiral arms is
almost lost in models with non-uniform $\ncr$, being far weaker than that
within 1--2\,kpc of the centre. Since we have truncated the angular velocity
at $r<2.1\kpc$, the untruncated differential rotation would lead to an even
stronger discrepancy.

Our synthetic maps do not include any depolarization effects due to random
magnetic fields (see Burn \cite{B66}; Sokoloff et al.\ \cite{S98}), although
they allow fully for depolarization by the regular magnetic fields
(differential Faraday rotation and beam depolarization). In order to
include Faraday depolarization effects due to the large-scale magnetic field,
we assumed a nominal constant ionization fraction of $X=\nel/n=0.1$,
corresponding to a thermal electron density of 0.1 of the total gas density
obtained from the gas dynamical simulations as described in Sect.~\ref{code}.
Guided by analogy with the Milky Way, where the average total gas density is
$1\cm^{-3}$ whereas the thermal electron density is $0.03\cm^{-3}$, a smaller
value of $X$ might be appropriate. We show results for $X$ close to this value
in Sect.~\ref{SynthPolMaps}.  In Sect.~\ref{ion_frac}, we discuss the effect of
variations in $X$ and argue that $0.01 \la X \la 0.2$.

We used two main techniques to compare the synthetic maps with observations and
therefore to select the optimal magnetic field model. We chose to use the
$\lambda6.2\cm$ map of polarized intensity in the analysis since it has the
best signal-to-noise ratio. All model data, including synthetic
radio maps, have been smoothed (in terms of the Stokes parameters $Q$ and $U$)
to match the resolution of the observations. In Sect.~\ref{Sect_cuts} we
compare the distributions of polarized intensity on cuts along various paths
in the plane of the sky. In Sect.~\ref{Difference_Maps} we analyse
the difference between the computed and observed polarized intensities in two
dimensions. In addition, 
we compare the orientations of the magnetic $B$-vectors obtained from
the observed and synthetic Stokes parameters (Sect.~\ref{orientation}).

To rotate the model galaxy to the position of NGC~1365 in the
sky, we took the inclination angle $i=46^{\circ}$ and the position angle
of the galaxy's major axis (i.e the intersection of the sky plane and the galaxy plane) $\PA=222^{\circ}$, which are those assumed
in obtaining the rotation curve for our (favoured) gas dynamical model.
Results are quite sensitive to these values, and it is possible that
a reappraisal could result in noticeable changes.

\begin{figure}\includegraphics[width=0.485\textwidth]{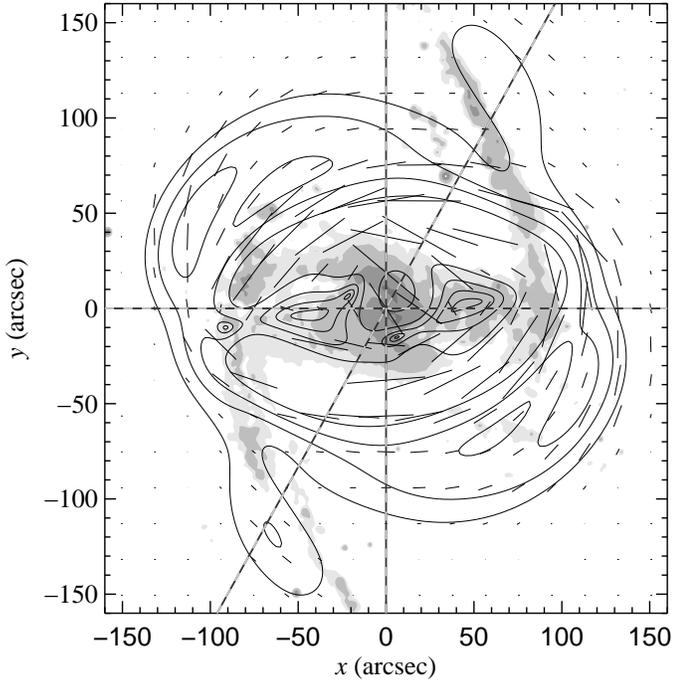}
\caption{
\label{synth_PI_map}
A synthetic map of polarized synchrotron intensity (contours) and polarization
planes at $\lambda6.2\cm$, resulting from
Model~2 (see Table~\ref{run_params})
assuming that $\ncr=\mbox{const}$, are shown superimposed on the optical
image of the galaxy NGC~1365 (shown in only a few shades of grey for
clarity). The synthetic
map has been smoothed to the resolution of $25\arcsec$ to match that of
the observed map shown in
Fig.~\ref{PIobs}. The contour levels shown are approximately
$(1, 3, 6, 12, 32) \times \PI_{\max}/ 45$, where $\PI_{\max}$ is
the maximum of $\PI$ in the synthetic map. Dashed lines show the
position of cuts discussed in Sect.~\ref{Sect_cuts}.
}
\end{figure}

\begin{figure}
\centerline{\includegraphics[scale=0.6]{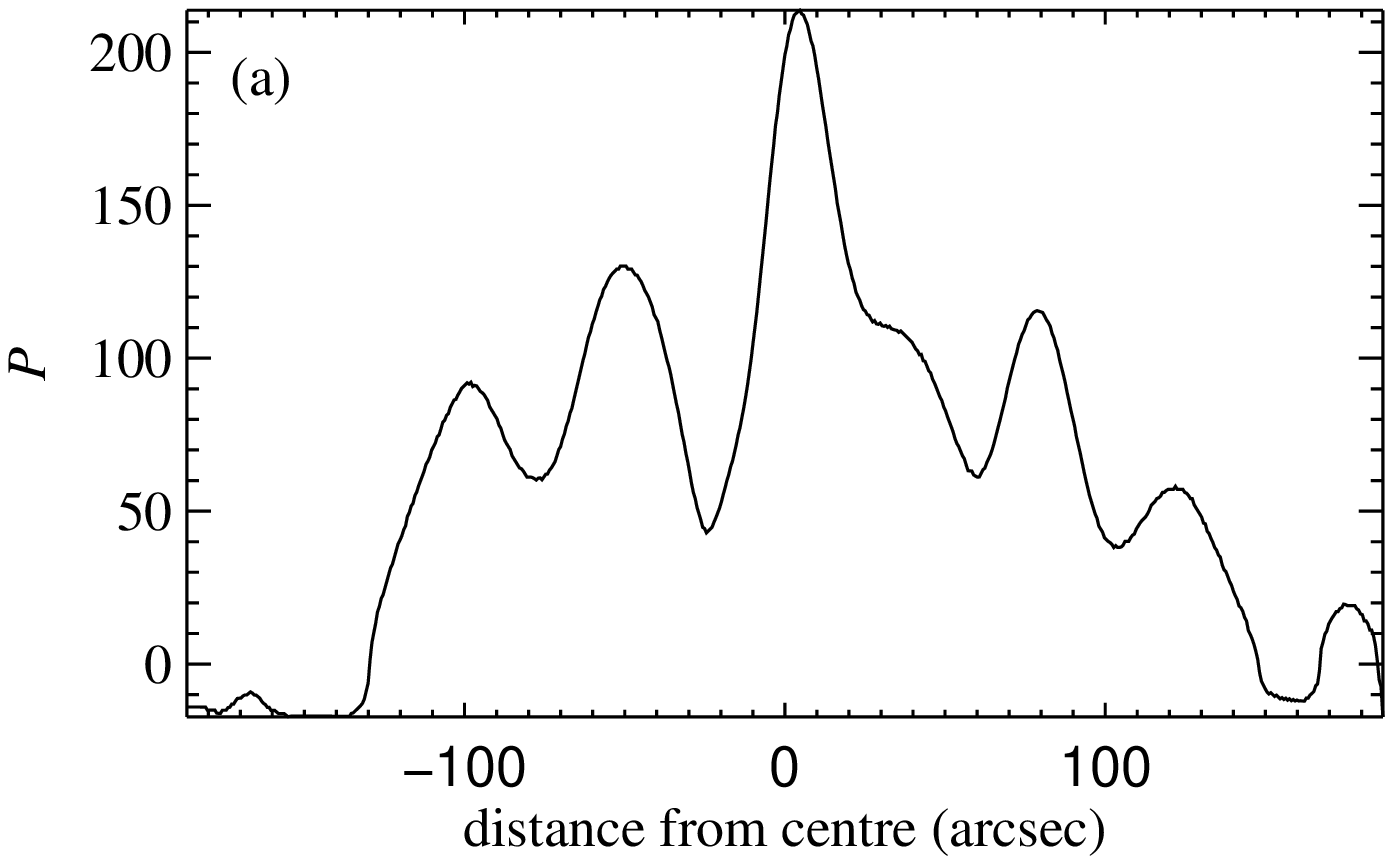}}
\centerline{\includegraphics[scale=0.6]{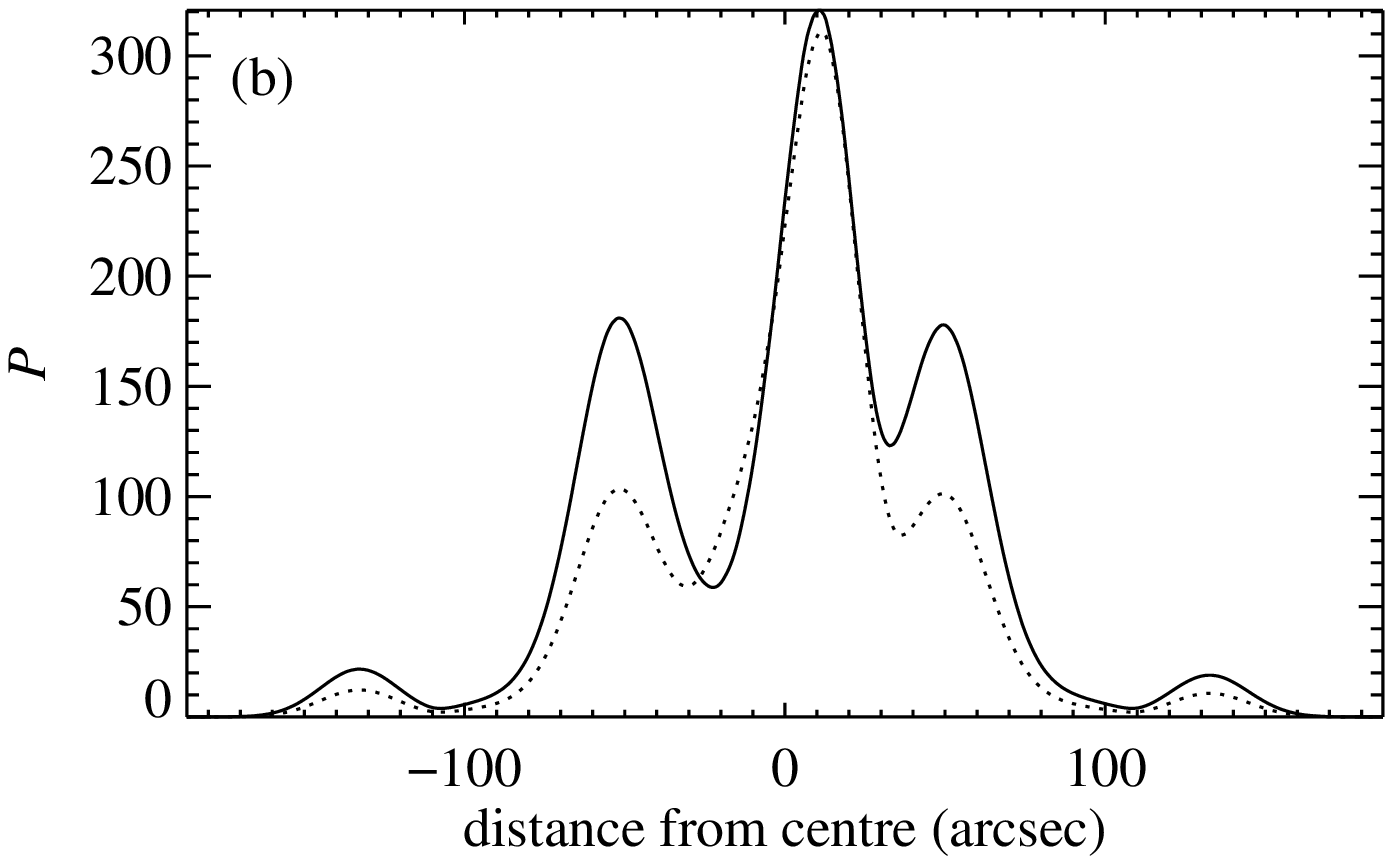}}
\centerline{\includegraphics[scale=0.6]{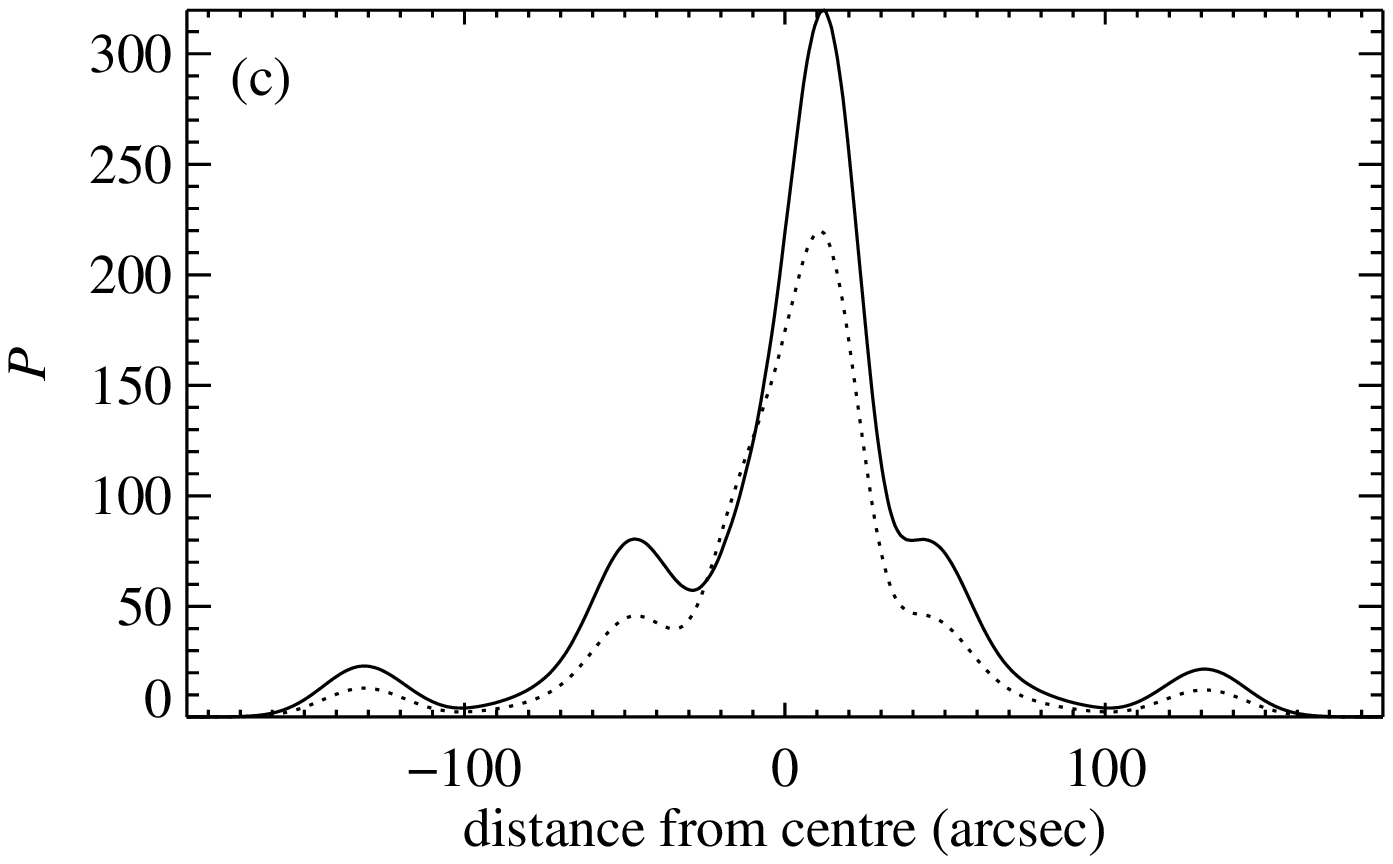}}
\caption{\label{ThirtyDegCuts}Cuts, at position angle $-31\degr$ passing
through the galactic centre (left to right in the plots corresponds moving from
south-east to north-west in the sky), through polarized intensity maps at
$\lambda6.2\cm$ smoothed to
$\mbox{HPBW}=25\arcsec$, for {\bf(a)} the observed map, and synthetic maps
from {\bf(b)} Model~2 and {\bf(c)} Model~4,
both for $\ncr=\mbox{const}$. In panels (b) and (c),
the synthetic profiles for $\lambda6.2 \cm$ and $\lambda3.5 \cm$ are shown solid and
dotted, respectively; the difference is due to Faraday and beam depolarization for the
assumed ionization degree $X=0.1$. The units of $\PI$ are as in
Fig.~\ref{PIobs} for (a)
and arbitrary in (b) and (c), but adjusted to fit a similar range.
The dotted profiles for $\lambda3.5 \cm$ with $X=0.1$ also correspond to $\PI$
at $\lambda6.2\cm$ with $X=0.032$.
}
\end{figure}

\begin{figure}
\centerline{\includegraphics[scale=0.6]{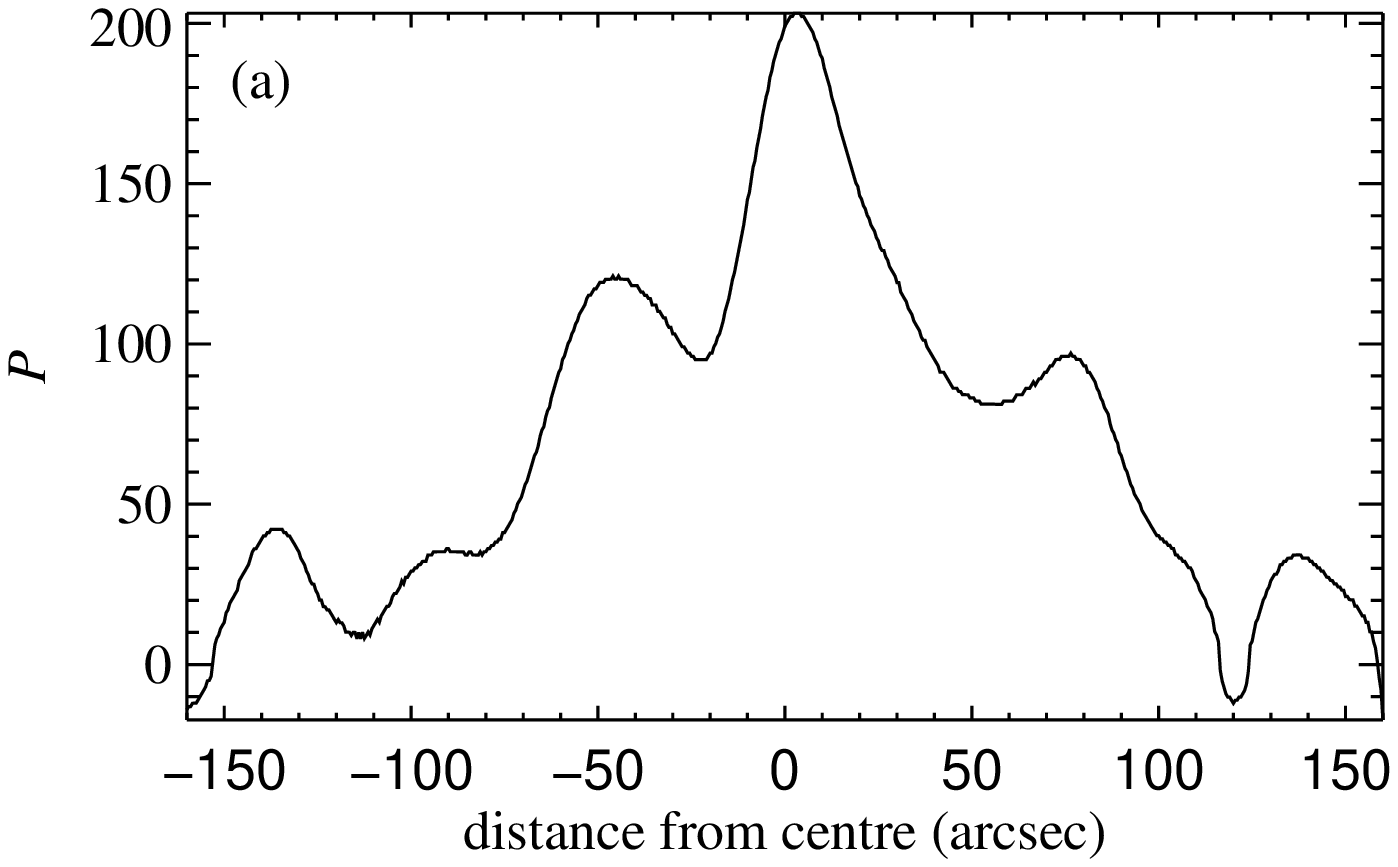}}
\centerline{\includegraphics[scale=0.6]{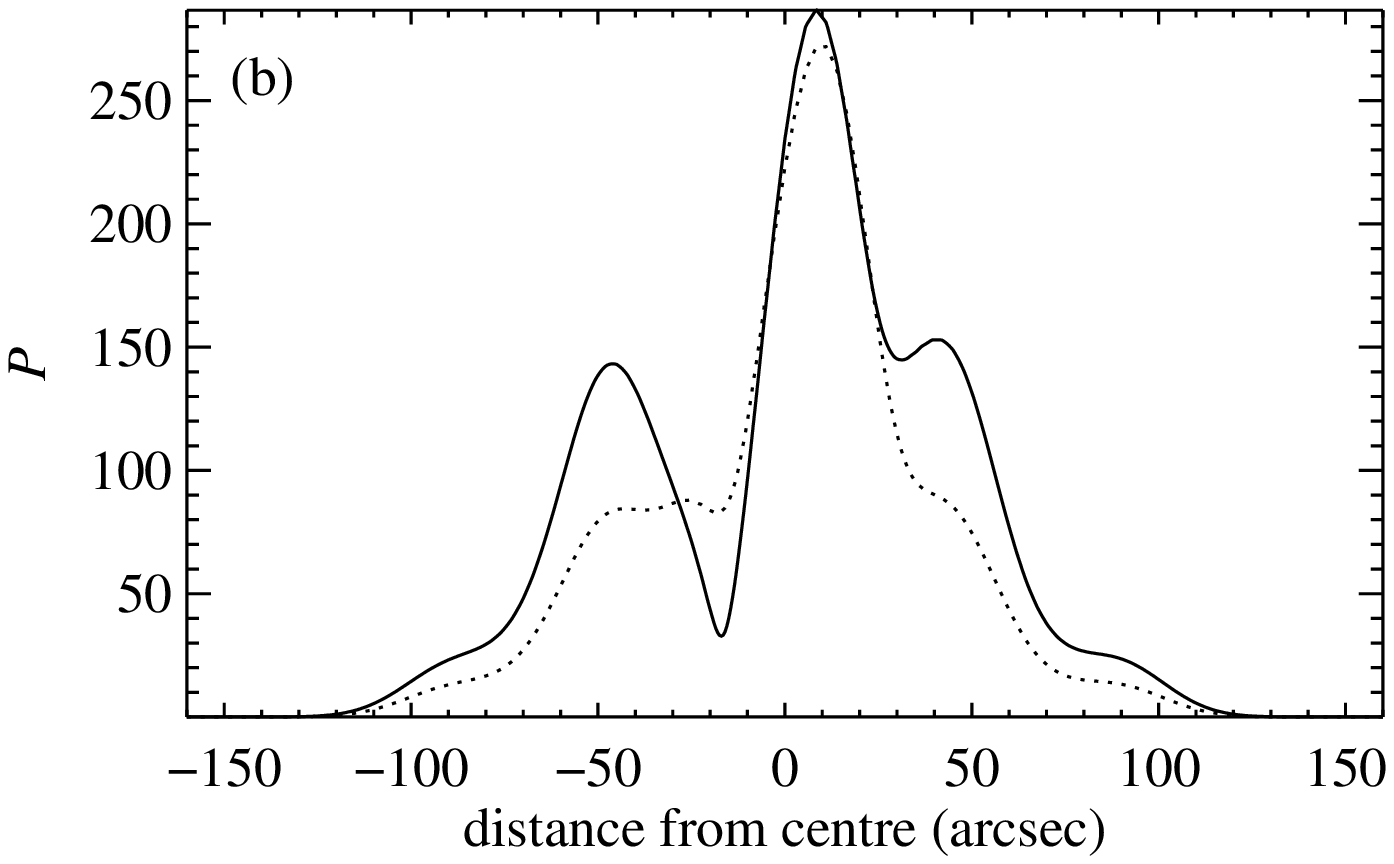}}
\centerline{\includegraphics[scale=0.6]{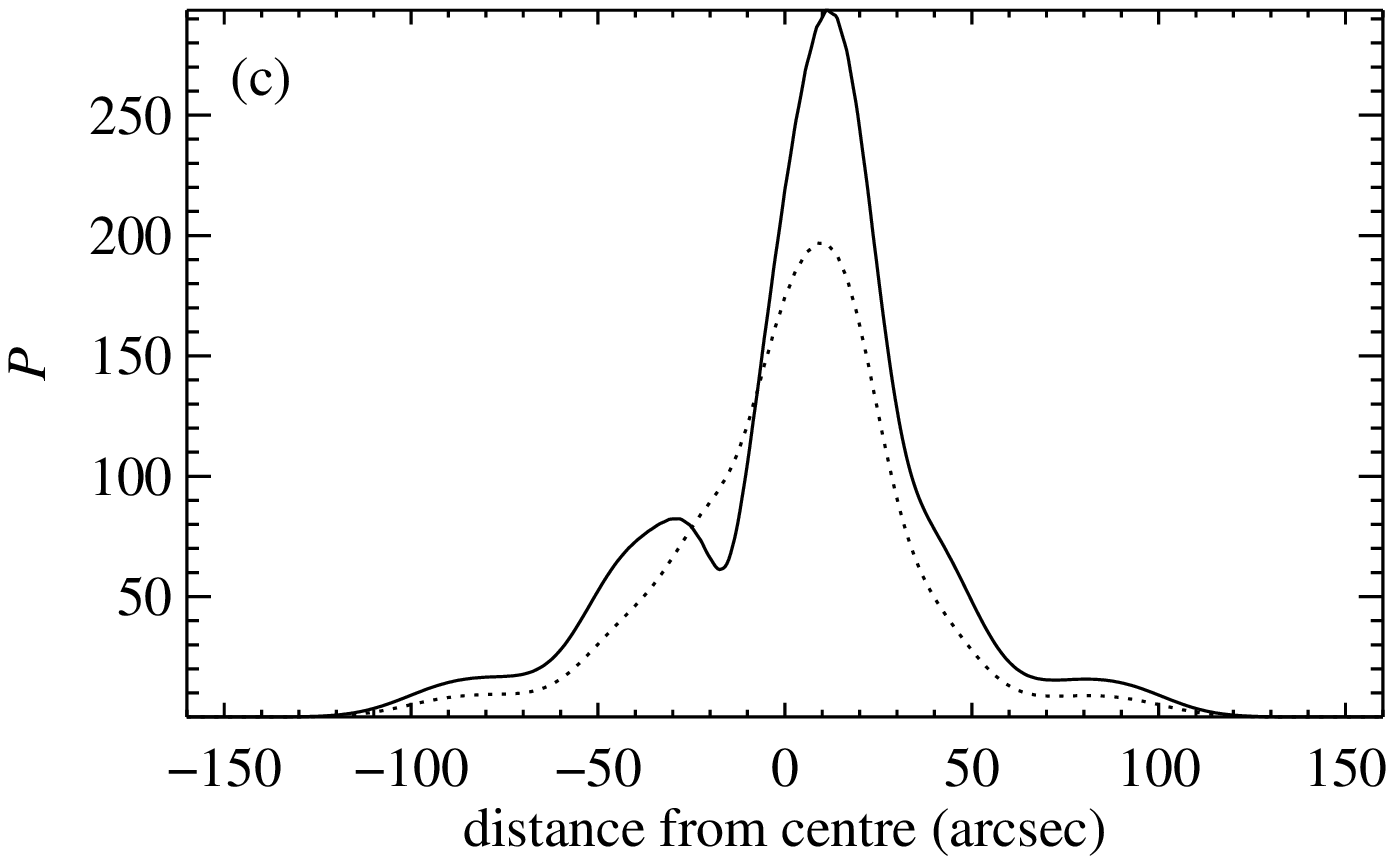}}
\caption{\label{ZeroDegCuts}As in Fig.~\ref{ThirtyDegCuts}, but at position
angle $0\degr$ (left to right is south to north in the sky).}
\end{figure}

\begin{figure}
\centerline{\includegraphics[scale=0.6]{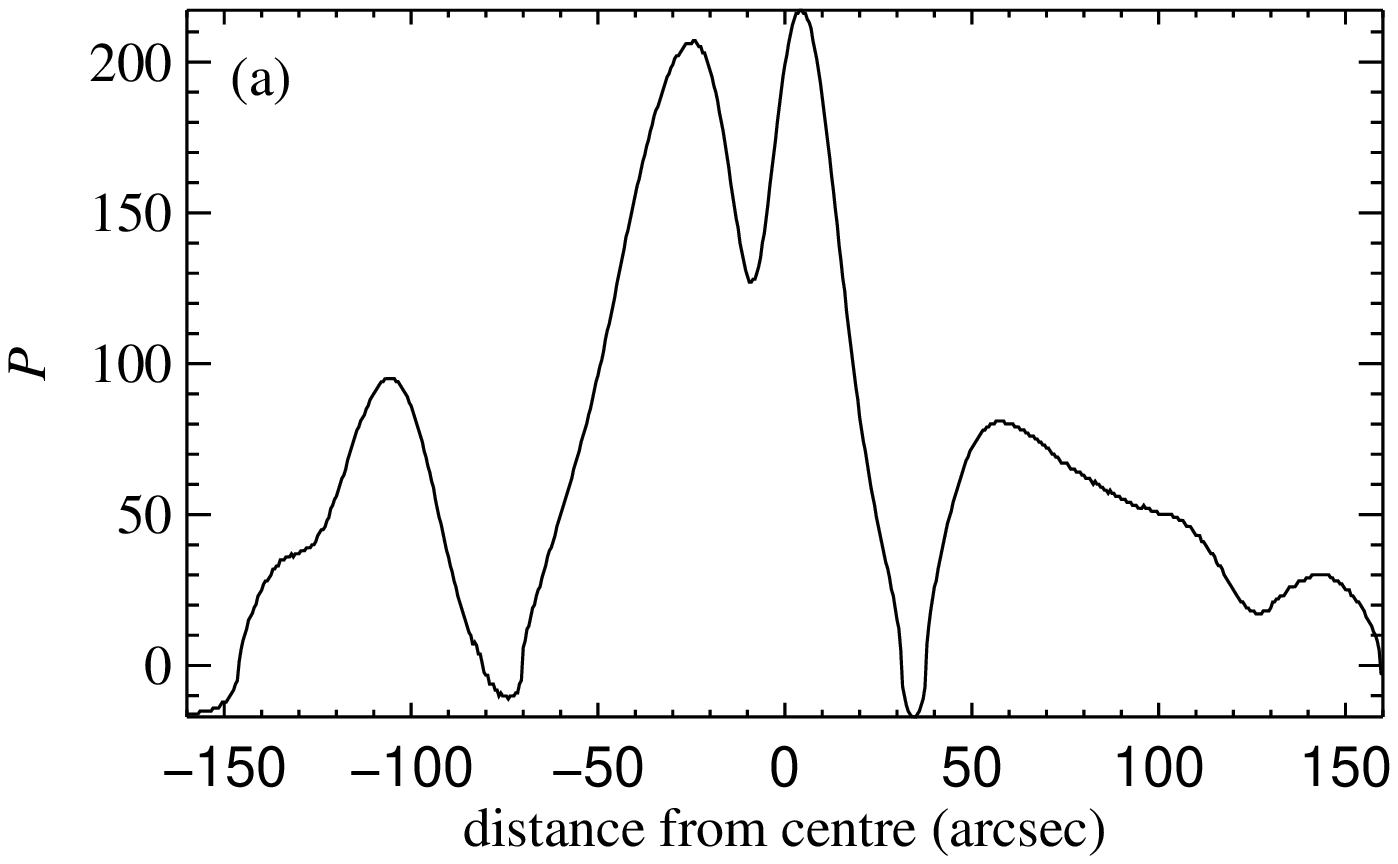}}
\centerline{\includegraphics[scale=0.6]{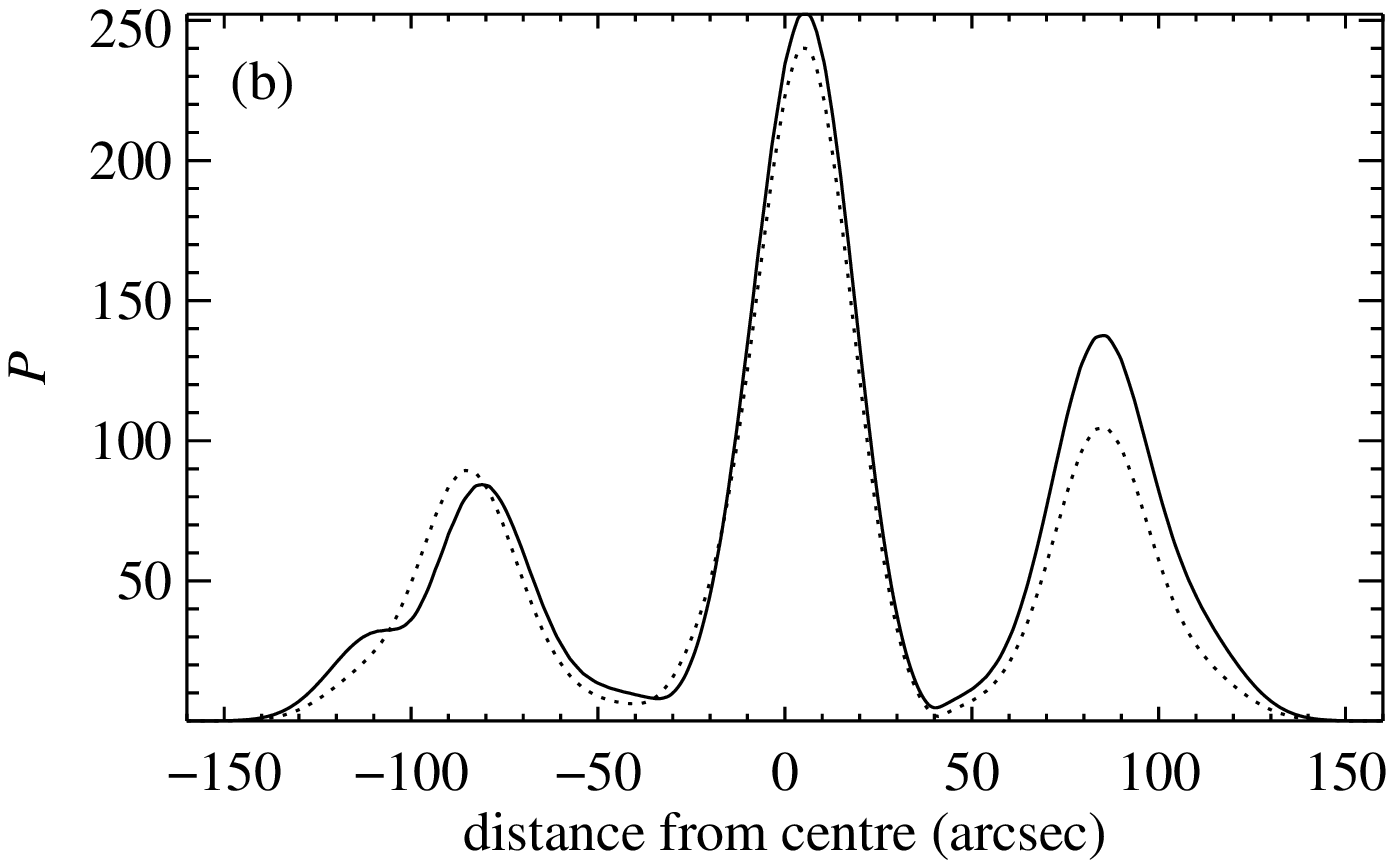}}
\centerline{\includegraphics[scale=0.6]{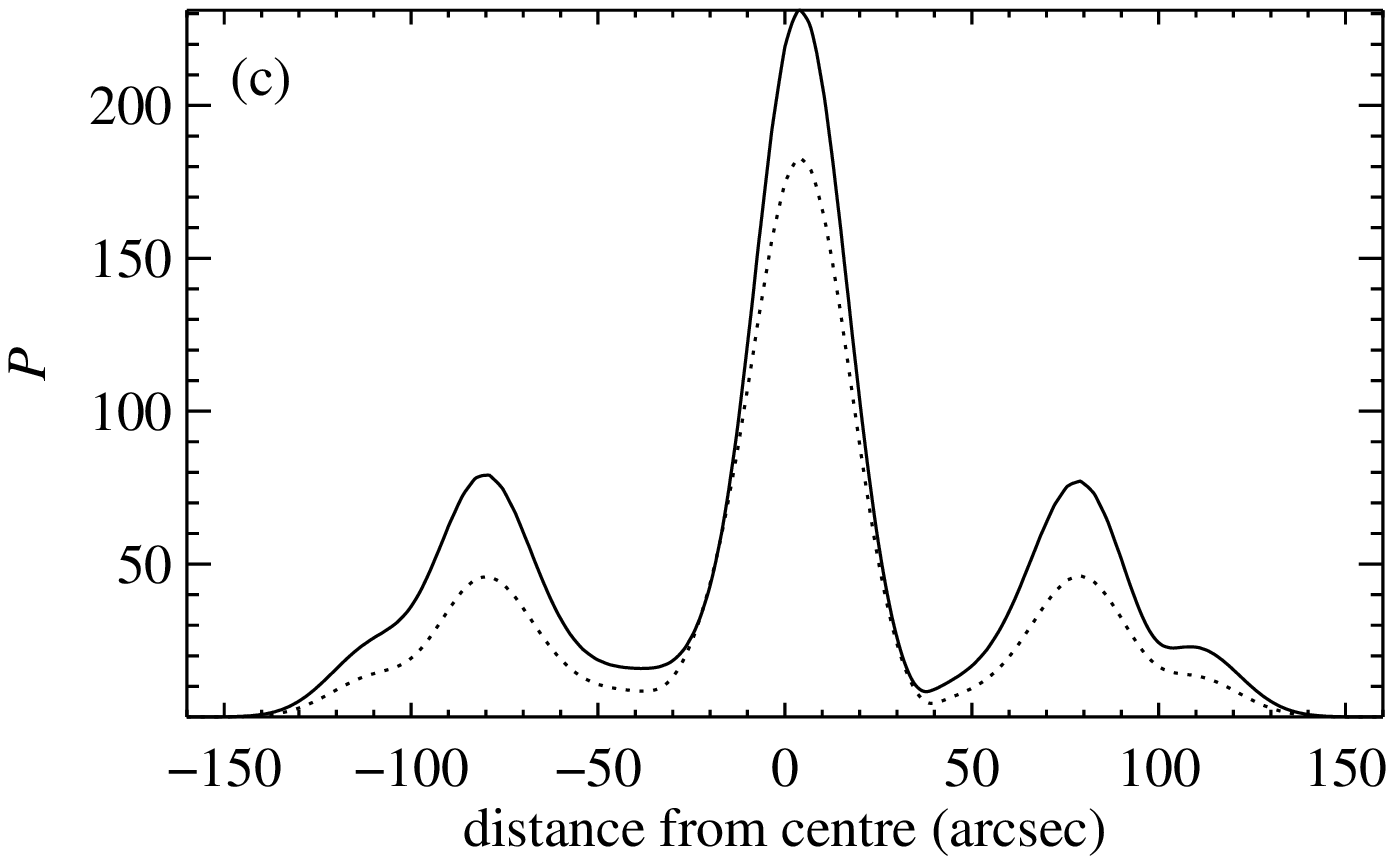}}
\caption{\label{NinetyDegCuts}As in Fig.~\ref{ThirtyDegCuts}, but at position
angle $-90\degr$ (left to right is east to west in the sky).}
\end{figure}

\subsection{Synthetic polarization maps}
\label{SynthPolMaps}
Overall, Model~2 (specified in Table~\ref{run_params}) appears to provide the best
fit to
the observed polarization map; Model~4 is only slightly worse -- see
Sect.~\ref{Sect_cuts}. Contours of $B^2$ shown in Fig.~\ref{B2} indicate that
the regular magnetic field is stronger in the bar region where gas density is
large, and outside the regions of high density in the spiral arms.
There are magnetic features apparently unrelated to the density distribution
[e.g., those passing through the positions $(x,y)\approx(-5,8), (5,-8)$]; they are
presumably formed by a locally enhanced velocity shear.
The magnetic field has a deep minimum within the bar, mainly produced by the
density deficiency in that region.
Other important features clearly visible in Fig.~\ref{B2} are
the magnetic field enhancements in the dust lane region, where magnetic field
is amplified by both compression and shear, and the prominent central peak.

The synthetic polarization map for this model is shown in Fig.~\ref{synth_PI_map}.
This can be compared directly with the observed map in the right-hand-panel of
Fig.~\ref{PIobs}; the maps (and all other maps we show) are at a similar scale to
facilitate the comparison; we make this comparison more quantitatively in
Sect.~\ref{Difference_Maps}. Our models have a high degree of symmetry, whereas the
`real' NGC~1365 is only approximately symmetric; since the observed map looks more
regular on the eastern side, we shall mostly refer to that part of the galaxy unless
stated otherwise. Despite the difference in symmetry, there is broad agreement
between these two maps; for example, both have a deep minimum of $\PI$ near the bar's
major axis where gas density is low, and both have the magnetic spiral arms displaced
from the gaseous ones (although both magnetic arms are displaced to larger
radii in the synthetic map, only one arm is so displaced in the observed map). The
minimum of
the synthetic $\PI$ in the bar (corresponding
also to a minimum of magnetic field within the bar, as seen in Fig.~\ref{B2}),
is broader than of the observations (see Sect.~\ref{Sect_cuts}).
The reason for this is the very low gas density in this
region, leading to weaker magnetic fields via Eq.~(\ref{alpha}). This feature
is further discussed in Sect.~\ref{Disc} where we argue that the gas dynamical
model underestimates significantly the amount of molecular gas in the bar
region.
Synthetic $\PI$ is
large both to the north and south of the bar major axis. In particular, the model
reproduces a maximum of $\PI$ upstream of the bar major axis, centred in the
$\lambda6.2\cm$ map of Fig.~\ref{PIobs} at $(\RA=03\mathrm{h}\,33\min\,40\sec,
\Dec=-36\degr09\arcmin00\arcsec)$. These maxima apparently arise from slightly enhanced
velocity shear (that locally amplifies magnetic field) rather than from local density
maxima. We also note maxima of $\PI$ near the ends of the bar and the beginning of the
spiral arms, at $(\RA=03\mathrm{h}\,33\,\min45\sec, \Dec=-36\degr08\arcmin15\arcsec)$
and $(\RA=03\mathrm{h}\,33\min\,28\sec, \Dec=-36\degr08\arcmin30\arcsec)$.
We note that the observed total emission (not shown here; see Beck et
al.~\cite{Betal05}) is related to gas density in a rather straightforward
manner being correlated with the gas density. The fact that this is not the
case with the polarized intensity (as seen in both observed and synthetic maps)
confirms that the observed regular magnetic field is not frozen into the
gas, apparently being affected by the dynamo action.

\subsection{Cuts through polarization maps}\label{Sect_cuts}
We found that comparisons can be usefully quantified and detailed using
the cuts in the sky plane mentioned above.
We show cuts only through the map at $\lambda6.2\cm$ because this map has
higher signal-to-noise ratio and includes the large-scale emission fully.
However, depolarization is significant at this wavelength (see Sect.~\ref{observations}) and
has to be taken into account when comparing the model and observations.
We use cuts through the centre of the
galaxy at position angles $\PA=0^{\circ},\ -90^{\circ}$ and $-31^{\circ}$,
where $\PA$ is measured counterclockwise from the north as shown in
Fig.~\ref{synth_PI_map}. (The angle $-31^{\circ}$ is chosen so that the cut
goes through the spiral arms; this corresponds roughly to a diagonal
in the computational frame of Fig.~\ref{B2}.) The
positions of these cuts are shown in Fig.~\ref{synth_PI_map}.
The synthetic $\PI$ has been normalized to make the mean difference
between that and the observed $\PI$ approximately zero.
We have superimposed another profile from cuts through synthetic maps
which represents both $\PI$ at
$\lambda3.5\cm$ with our favoured value of $X=0.1$, and also at $\lambda6.2\cm$
with $X=0.032$ (incidentally, this is close to the mean ionization degree of
the warm diffuse gas in the Milky Way). The coincidence of these two cuts is
due to equivalent
Faraday depolarization, which depends directly on the quantity
$ {\psi(z)} \propto {\lambda^{2} X \int_{z}^{\infty}n B_{\parallel} dz}$
along the line of sight towards an observer
(see Appendix~\ref{synthetic_radio_maps}).
Here $B_{\parallel}$ is the line of sight field component and $n_{\mathrm e} = X n$,
therefore  $\lambda^{2}X = \mbox{const}$ identifies equivalence in depolarization.
We see that this value is about the same in both cases (i.e.
$6.2^{2}\times 0.032 \approx 3.5^{2}\times 0.1$). The difference
between the polarization for these two possibilities is then just a
$\lambda$-dependent scale factor.

The cuts are presented in Figs~\ref{ThirtyDegCuts},
\ref{ZeroDegCuts} and \ref{NinetyDegCuts} for the best-fit Model~2 and also for
Model~4. The latter model has the background turbulent magnetic diffusivity
$\eta_0$ enhanced by a factor of 2. This leads to a significantly smoother,
less structured
distribution of $\PI$. Thus, comparison of Models 2 and 4 allows us to
suggest that the effective
turbulent magnetic diffusivity in the interstellar gas of barred galaxies is,
on average,
close to $\eta_0=10^{26}\cm^2\s^{-1}$. This value is typical of spiral
galaxies in general and is that obtained if the turbulent speed $v_{\rm t}$
is close to $10\kms$ and the turbulent scale is about
$l=0.1\kpc$; $\eta_0\simeq\frac13 lv_\mathrm{t}$.

Our model neglects depolarization due to random magnetic fields which can
reduce the value of $\PI$ in the central parts more strongly than in the outer
galaxy and therefore affect the relative height of the central peaks in
Figs~\ref{ThirtyDegCuts}--\ref{NinetyDegCuts}. Depolarization due to
internal Faraday dispersion reduces the degree of polarization to
\begin{equation}\label{depol}
p=p_0\frac{1-e^{-S}}{S}\;,
\end{equation}
where $S=2\sigma_\RM^2\lambda^4$ with $\sigma_\RM^2=2C_1^2\langle
b^2\rangle\langle\nel^2\rangle dL$ the variance of the Faraday rotation
measure. Here $C_1$ is the dimensional constant appearing in the definition of
the Faraday rotation measure (see Appendix~\ref{synthetic_radio_maps}), $b$ is
the turbulent magnetic field, angular brackets denote averaging (the
fluctuations in magnetic field and thermal electron density are assumed to be
uncorrelated), $d$ is the turbulent scale and $L$ is the path length (Sokoloff
et al.\ \cite{S98}). The best available estimate of the random magnetic field
in the central region of NGC~1365, $b\simeq40\mkG$, follows from the total synchrotron intensity
assuming equipartition between cosmic rays and magnetic fields (see however
Sect.~\ref{n_cr_models} for a discussion of the validity of this assumption).
For $\nel=0.03\cm^{-3}$, $d=0.1\kpc$, $L=1\kpc$
and $\lambda=6.2\cm$, we then obtain $S\simeq6$, implying that this mechanism can
depolarize the central peak significantly, giving $p/p_0\simeq0.2$.
Since the height of the secondary peak should also be affected by depolarization,
albeit to a lesser extent,
we expect that the ratio of the two peaks will be reduced by a factor smaller
than five. We note, however, that this
estimate is uncertain since the  number density of
thermal electrons, their filling factor, turbulent scale and other parameters
are not known well enough. An alternative is to assess the importance of
this depolarization effect by comparing polarized intensities at $\lambda6.2\cm$
and $\lambda3.5\cm$. The ratio of the central peak to the secondary ones
at $\lambda3.5\cm$ is about 6--8, as opposed to 2--3 at $\lambda6.2\cm$. The
difference can be attributed to Faraday depolarization (by both regular and random
magnetic fields). Assuming that Faraday depolarization at $\lambda3.5\cm$ is
negligible, we conclude that it can reduce the degree of polarization at
$\lambda6.2\cm$ by a factor as large as 4, which is consistent with the
analytical estimate. We conclude that the relative height of the central
peak in the synthetic cuts of Figs~\ref{ThirtyDegCuts}--\ref{NinetyDegCuts}
would be reduced by Faraday dispersion, although this is difficult to estimate
accurately.

Given the above uncertainties in the amount of depolarization, all three
cuts for Model~2 are similar to those observed. In particular, the relative heights of
the peaks in $\PI$ and, more importantly, the positions of both maxima and minima
are remarkably realistic. The characteristic feature of this model is that $\eta$ is
further enhanced by a factor of $q_\eta\simeq3$ in the inner region of NGC~1365,
$r\la3\kpc$. This enhancement can be due to a higher rate of star formation, and
hence more hot gas, with a correspondingly higher speed of sound, which
would allow the turbulent speed to be larger than elsewhere.

The Model~2 cut at $\PA=-31\degr$ (Fig.~\ref{ThirtyDegCuts}), which passes through
the spiral arms, shows an encouraging agreement with observations.
For example, $B$ has a maximum slightly outside the northern
arm in both this model and the real galaxy. However, the outermost maxima
produced by the spiral arms are slightly too far away from the centre in the
model. As illustrated in Fig.~\ref{ThirtyDegCuts}(b), the relative heights of the
peaks at $\lambda6.2\cm$ are significantly affected by Faraday rotation even for
$X=0.1$, where they clearly differ by more than just a scale factor between
$\lambda6.2 \cm$ and $\lambda3.5 \cm$.

The cut at $\PA=0$ (Fig.~\ref{ZeroDegCuts}) exhibits similar degree of agreement
with the observations. The main deficiency of the model here is the
too narrow distribution of $\PI$ (the magnetic structure of the model is too poor
outside the bar) and the minimum is too deep near the centre of the cut.

The cut at $\PA=-90\degr$ in the synthetic map, shown in
Fig.~\ref{NinetyDegCuts}, has a central maximum that is too narrow
(or off-centre minima that are too broad).
This difference results in the deep minima in the difference
parameter $\delta$ discussed in Sect.~\ref{Difference_Maps}.
The sharp minimum in the observed cut near the centre is a result
of beam depolarization; it occurs in the synthetic cuts as well, but is
removed by smoothing.

Model~2 seems to be almost optimal. The model could be fine tuned by changing
$\eta_{0}$ and $r_{\eta}$ within the ranges (1--$2)\times10^{26}\cm^2\s^{-1}$
and 1.5--3, respectively. For example, the secondary peaks in the $\PA=0$ cut
decrease in strength in Model~4. Further, increasing $\ncr$ by a factor of 2
within the central 1.5\,kpc would make the central peak higher. However, we
have not made such {\it post hoc\/} adjustments.

\begin{figure*}
\begin{center}
\includegraphics[width=0.7\textwidth]{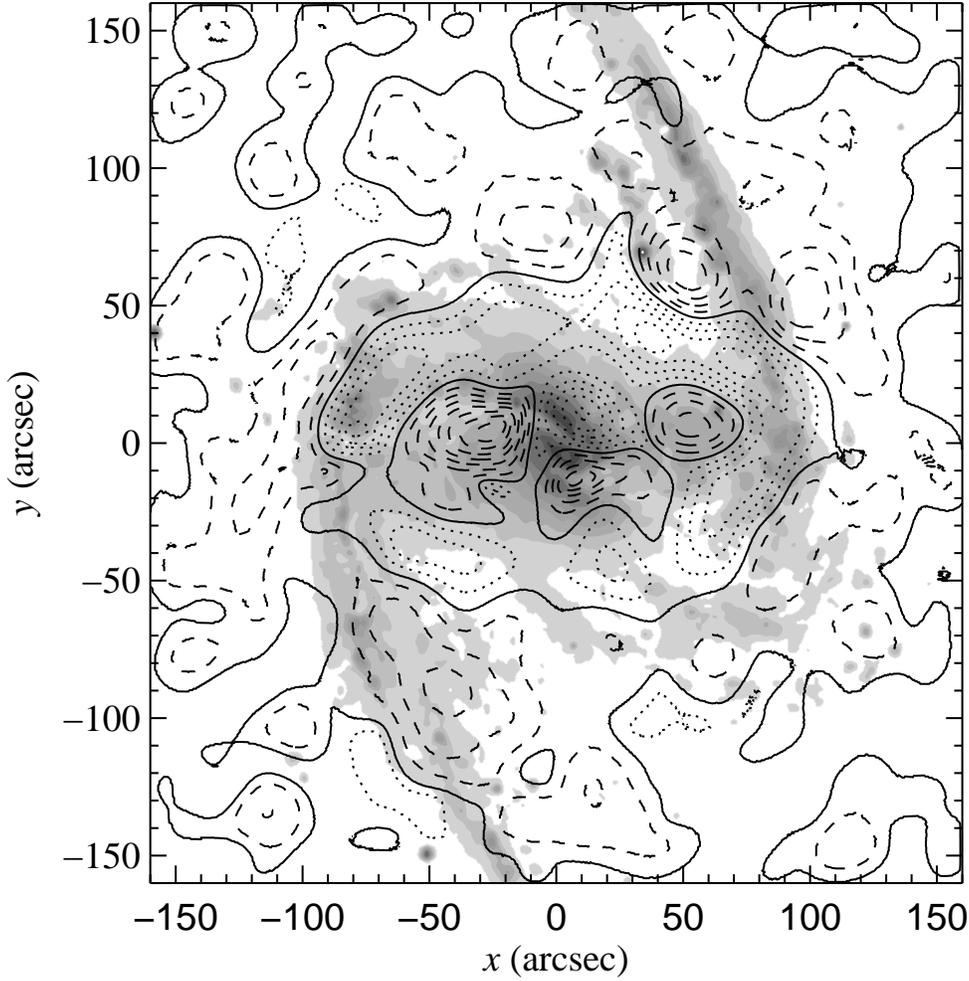}
\end{center}
\caption{\label{diff_map}The difference $\delta$ between normalized
synthetic and observed polarization maps at $\lambda6.2\cm$, as defined in
Eq.~(\ref{del}), superimposed on the optical image of NGC~1365. The contour
spacing is 2, with the zero contour
shown solid, negative values of $\delta$ dashed,
and positive, dotted.}
\end{figure*}

\subsection{The difference maps} \label{Difference_Maps}
To obtain a global comparison of the models and observations, we produced maps
of the difference between the observed and synthetic polarization
at $\lambda6.2\cm$, with the synthetic polarization scaled to make the mean difference
approximately zero;
this measure was further normalized by dividing
the difference by the appropriately normalized noise level of
the observed map giving
\begin{equation}\label{del}
\delta=
\frac{(1.4\PI/\PI_\mathrm{max})_\mathrm{model}
-(\PI/\PI_\mathrm{max})_\mathrm{obs}}{(\sigma_\PI/\PI_\mathrm{max})_\mathrm{obs}}\;.
\end{equation}
Thus, all comparisons were performed pointwise
after their reduction to the common resolution $25\arcsec$ -- this is quite a stringent
test of the model.
The result is shown in Fig.~\ref{diff_map} for Model~2.

Since the models --
unlike the real galaxy -- possess perfect symmetry, the difference can hardly
be uniformly small: a perfect fit in one half of the galaxy would produce
significant systematic discrepancy in the other half. With this caveat, the
difference map shows an acceptable global agreement of the model with
observations, in that it does not show much of the basic morphological elements of the
galaxy. The normalized relative difference is about 6--14 in four spots observed to
the east, south and
north-west of the galactic centre, indicating that synthetic polarized intensity is
too small
upstream of the dust lanes and at two positions at the inner edge of the
western spiral arm. Otherwise, $|\delta|\la4$ across the whole field of view. Given
the limited scope of our model (e.g., it does not include
any turbulent magnetic fields which can produce polarized radio emission where
they are anisotropic), we consider this degree of agreement to be acceptable.
We discuss in Sect.~\ref{n_cr_models} a cosmic ray distribution that would
provide perfect fit of Model 2 to observations.



\subsection{Faraday rotation}
\label{Faraday}

We can use polarized intensity (as in the comparisons above) to probe the
distribution of the large-scale magnetic field strength, and also to deduce the
orientation of the magnetic field in the plane of the sky (via polarization vectors).
However, knowledge of this quantity does not determine the field direction.
The Faraday rotation measure $\RM$ is sensitive to the
direction of the magnetic field, but the observed $\RM$ map is very patchy
because of the lower signal-to-noise ratio at $\lambda3.5\cm$. Therefore, we
used $\RM$ data only to establish a minimum acceptable degree of gas ionization.



\subsection{Magnetic field structure}
\label{orientation}

\begin{figure}
\includegraphics[width=0.47\textwidth]{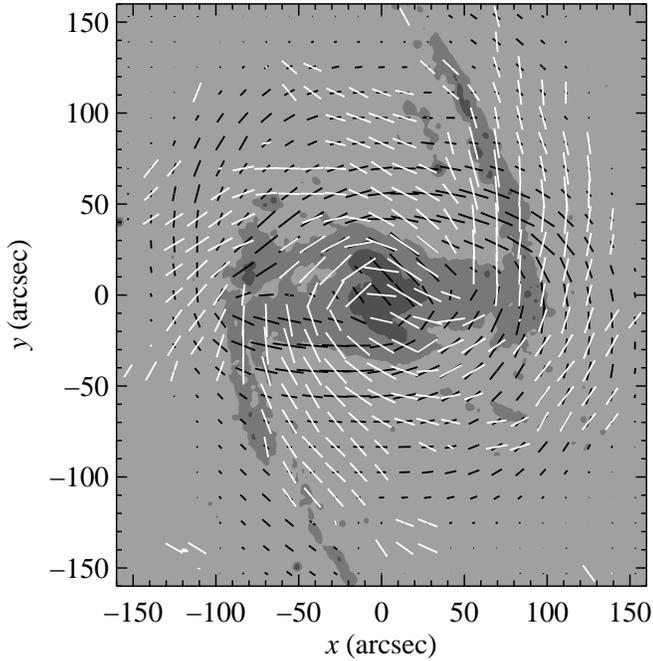}
\caption{Orientations of the $B$-vectors
of polarized emission obtained from the observed (white)
and synthetic (black, Model 2) Stokes parameters at $\lambda6.2\cm$.  The
vector lengths are proportional to $\sqrt{\PI}$. For the observed $B$-vectors,
positions where polarized intensity is weaker than three times the noise level,
$\PI < 3\sigma_\PI$, are neglected.
}
\label{Bvectors-uncorrected}
\end{figure}

An analysis of the observed global magnetic structure in NGC~1365
that is sensitive to the direction of magnetic field was performed by
Beck et al.\ (\cite{Betal05})
by fitting the polarization angles obtained from multi-frequency observations.
This analysis provides the large-scale magnetic field
expanded into Fourier series in the azimuthal angle.
Their results indicate the presence of
a significant component with the azimuthal wave number $m=1$ at almost all
distances from the galactic centre. However, our underlying gas dynamical model
has even symmetry in azimuth, so that modes with odd values of $m$ do not occur
in the modelled magnetic field.
The contribution of the $m=1$ mode to those Fourier expansions is more important
than just producing the overall asymmetry.
In particular, superposition of various azimuthal modes produces
local magnetic features at kiloparsec scale which are lost if only even modes are
retained in the observed structure to facilitate comparison with the model.
Therefore, we did not find it useful to compare the modelled and
observed magnetic structures in this manner.
(The presence of unmodelled odd-$m$ structure was also a feature
of our study of NGC 1097 in Moss et al.\ \cite{MS01}.)

We instead compare directly the orientation of the magnetic field vectors
in the observed and synthetic polarization maps. Comparison of 
two-dimensional vector fields is difficult. We could approach this by taking
cuts through maps of the magnetic field orientation angles, as was done with
the polarized intensity. However, a small shift in a feature such as a shock
front can result in drastic differences between any such cuts made parallel
to the front.

In Fig.~\ref{Bvectors-uncorrected} we show the orientation of both the
synthetic and observed magnetic field vectors obtained directly from the
corresponding Stokes parameters; points below $3$ times the r.m.s. noise
level are neglected in the observed maps.
Agreement between model and observations is reasonable in the top left and
bottom right quadrants near the bar (and partly further out), whereas the
differences are quite large in the other regions.  The difference  is especially large near beginning of the spiral arms. The mean value of the difference between the observed and modelled polarization angles is $33^\circ$, and its standard deviation is $25^\circ$. For comparison, the errors in the observed polarization angle range from 2 to 10 degrees.

The overall difference is that the model polarization vectors are arranged into a more
elliptical pattern around the bar than the observed ones, which exhibit a more circular configuration (and have almost constant pitch angles). 
It seems that the nonaxisymmetric distortion due to the bar is weaker in the observed magnetic field than in the model. This could be because the magnetic 
field is coupled to a warm gas component which has less response to the
bar's potential than cold gas and stars.
We made a comparison similar to that in Fig.~\ref{Bvectors-uncorrected} but for Model 6, where the speed of sound is $30\kms$ (see Table~\ref{run_params}). The resulting gas dynamical model illustrated in Fig.~\ref{gas_1030} (right hand panel) has a more uniform density distribution and weaker deviations from axial symmetry. The improvement in the magnetic pattern was only marginal, and so the reason for this discrepancy remains unclear.

%
\section{Sensitivity to parameters, and implications of the dynamo models}
In this section, we discuss how synthetic radio maps are affected by various
changes in our model. This allows us to infer useful information about
the interstellar medium in the galaxy.

\subsection{Distribution of cosmic rays}\label{n_cr_models}
\begin{figure}
\begin{center}
\includegraphics[scale=0.6]{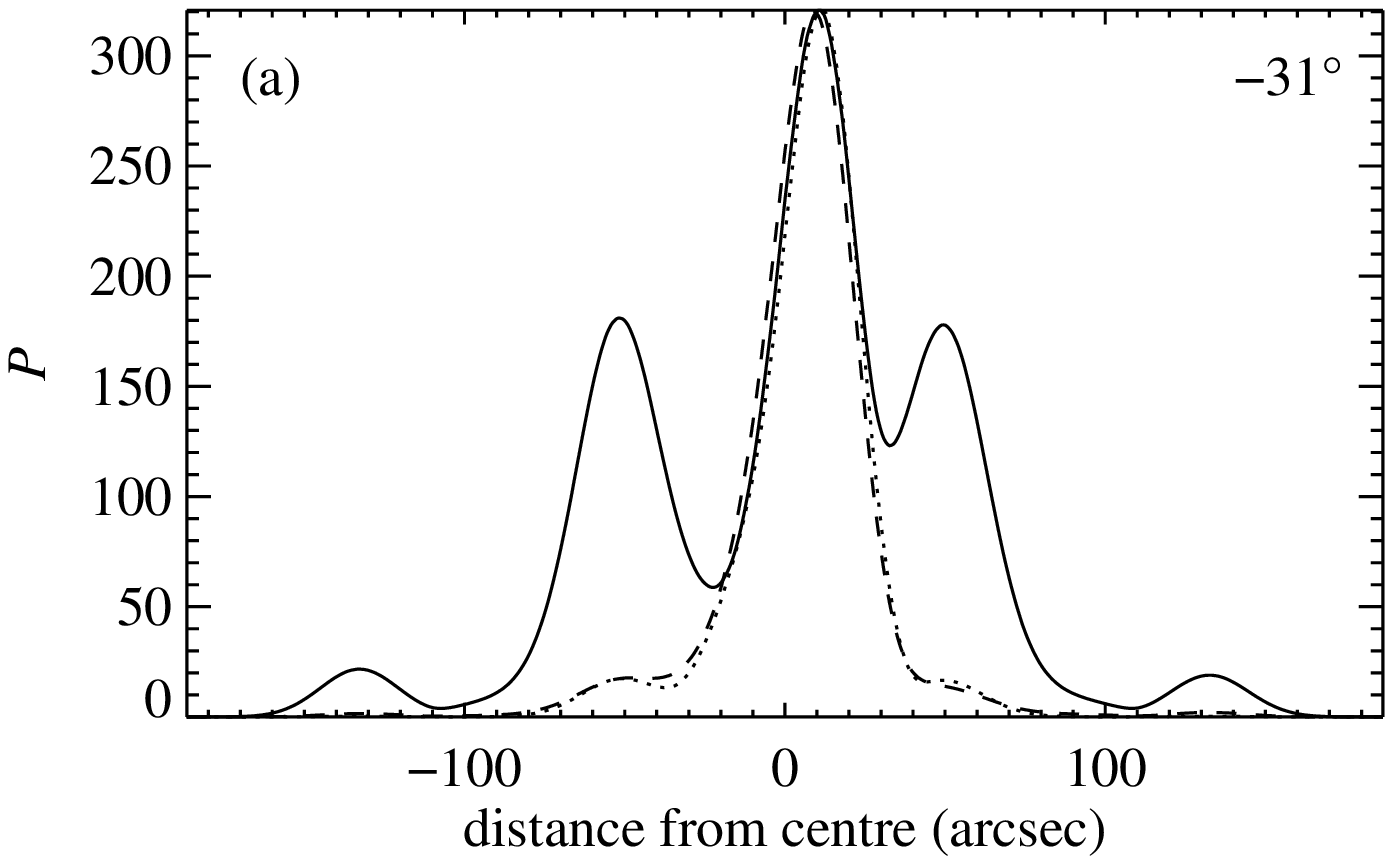}
\includegraphics[scale=0.6]{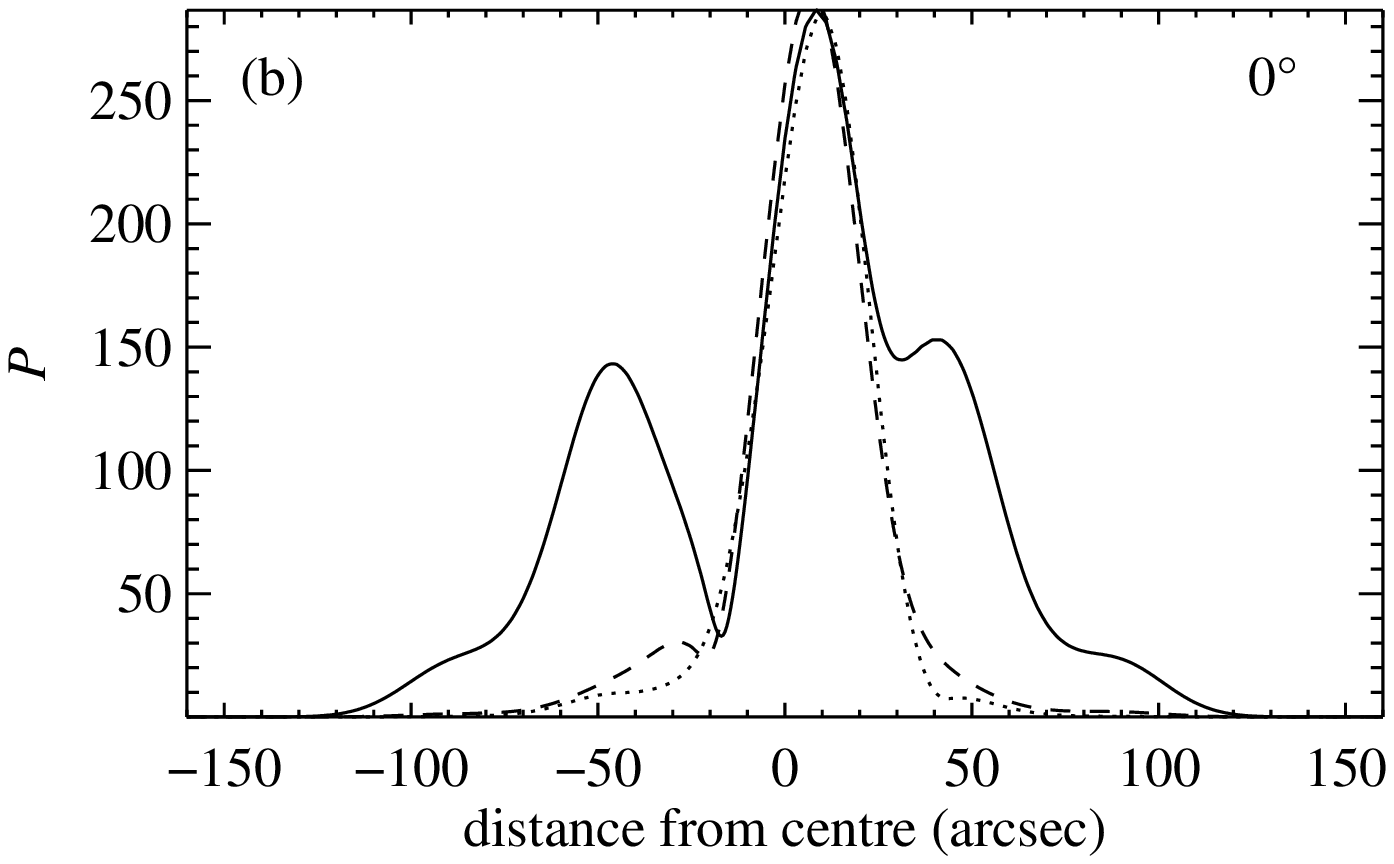}
\includegraphics[scale=0.6]{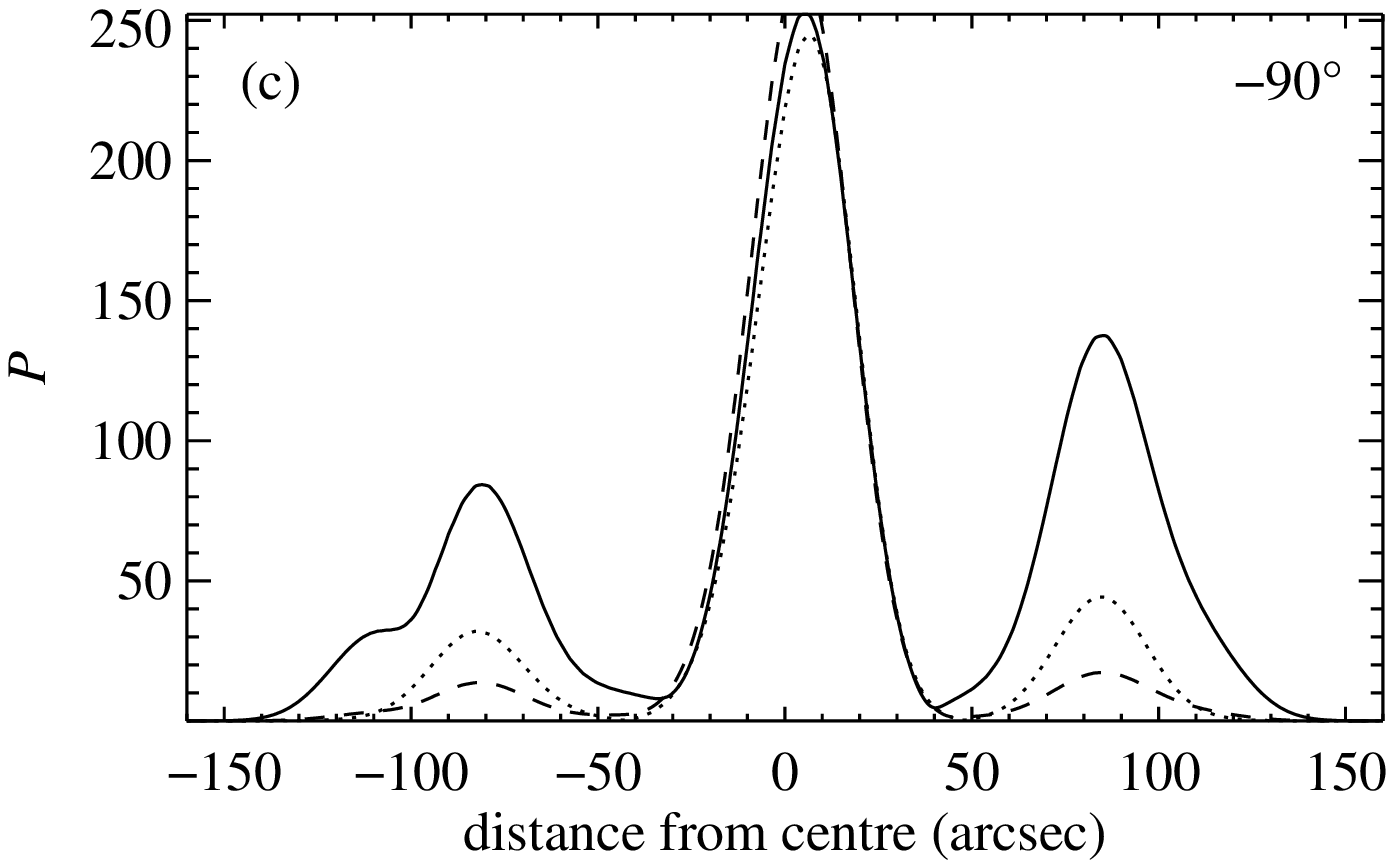}
\end{center}
\caption{\label{c_ray_effect}Cuts through radio maps for various models of
cosmic ray distribution as defined in Eq.~(\ref{crayd});
the position angle of the cut is indicated in the upper right corner of
each panel: {\bf(a)} $-31^\circ$,  {\bf(b)} 0, and {\bf(c)} $-90^\circ$. The solid lines
are for Model~2, i.e., with  $n_\mathrm{cr}=\mbox{const}$,
and are identical to those in
Figs.~\ref{ThirtyDegCuts}(b)--\ref{NinetyDegCuts}(b); the other two cuts in
each panel are for
model (ii) $n_\mathrm{cr}=\propto\sqrt{I}$, dashed; and model (iii)
$n_\mathrm{cr}=p^{-1}$ , dotted.
}
\end{figure}

In order to calculate the synchrotron intensity, we need to specify the number
density of cosmic ray electrons. There are no direct measurements of this
quantity in external galaxies, and there are no sufficiently detailed theories
that might supply it.
Therefore, empirical models for the cosmic ray number density $\ncr$ were
studied with
\begin{equation}\label{crayd}
\quad\begin{minipage}{0.2\textwidth}
\begin{enumerate}
\item[(i)] $\ncr=\mbox{const}$,
\item[(ii)] $\ncr\propto\sqrt{I}$,
\item[(iii)] $\ncr\propto p^{-1}$,
\end{enumerate}
\end{minipage}
\end{equation}
where $I$ is the observed total radio intensity and $p$ is
the percentage polarization of synchrotron emission at short wavelengths;
models (ii) and (iii) are motivated below.
Thus, we make an attempt to relate $\ncr$ to
various observable quantities.
The first model is the simplest possible, and
it attributes all the variation in synchrotron intensity to that of the
magnetic field. The motivation for model (ii) is that, if cosmic rays are in
equipartition with the total magnetic field, $\ncr\propto B_\mathrm{tot}^2$,
then the total intensity of synchrotron emission is
roughly proportional to $\ncr^2$.
Model (iii) relies again on the idea of equipartition between cosmic rays and
magnetic fields, but now we use expression $p\simeq p_0 B^2/B_\mathrm{tot}^2$
(Burn \cite{B66}, Sokoloff et al.\ \cite{S98}) to estimate $\ncr$ as
$\ncr\propto B_\mathrm{tot}^2\simeq B^2 p_0/p$, where $p_0\approx0.75$ and $B$
can be taken from our dynamo model.
This rough estimate neglects any depolarization effects and any
contribution of anisotropic random magnetic fields to polarized intensity
(Sokoloff et al.\ \cite{S98}).

As illustrated in Fig.~\ref{c_ray_effect}, the major and universal
effect of any plausible non-uniform distribution of $\ncr$ is to enhance the
central maximum in $\PI$, so that the peaks in the outer parts become relatively
insignificant. Most importantly, any signature of the spiral arms
almost disappears.
If the synthetic polarized intensity is rescaled to fit that observed
in the spiral arms, the central peak becomes unacceptably broad and high.
Using $n_\mathrm{cr}\propto B^2/p$ instead of $p^{-1}$ as in model (iii)
does not improve the situation. Similarly, it does not help if we use
$\ncr\propto B_\mathrm{tot}^2=B^2+4\pi\rho v_\mathrm{t}^2$, assuming that the
random magnetic field is in equipartition with turbulent energy.
Since models (ii) and (iii) involve the assumption of energy equipartition
(or pressure balance) between cosmic rays and magnetic fields, we conclude
that our results
do not suggest this type of relation between cosmic rays and magnetic fields at
large scales.

We can think of several plausible explanations
for this perhaps surprising result. It may be that some fraction
of the polarized emission in the outer galaxy is produced by anisotropic
magnetic fields which are not modelled. We do not consider this to be a very
plausible option as this would require
that the anisotropy is larger in the outer galaxy and in the spiral arms, rather
than in the region of the central peak. Meanwhile, velocity
shear, which might produce the anisotropy, is stronger in the inner bar region.
More plausibly, cosmic ray diffusion
makes their distribution smoother than that of the magnetic field. With the
cosmic ray diffusivity of $K\simeq10^{29}\cm^2\s^{-1}$ and the confinement time
$\tau\simeq10^6\yr$, their distribution would be rather homogeneous at scales
$(K\tau)^{1/2}\simeq1\kpc$. We note, however, that our model suggests that cosmic ray
distribution is almost uniform at scales of order 10\,kpc.

It cannot be excluded that the synthetic polarization maps exaggerate
the relative height of the central peak because they neglect Faraday depolarization
due to random magnetic fields, namely the internal Faraday dispersion discussed
in Sect.~\ref{Sect_cuts}. As follows from Eq.~(\ref{depol}), and discussion
following it, it is not implausible that this effect can reduce the relative
height of the central peak by a factor of five or somewhat less.
As can be seen from Fig.~\ref{c_ray_effect}, the ratio of the central to the
secondary peaks at a distance of about $50\arcsec$ from the centre is about
20 or more for the non-uniform distributions of cosmic
rays, whereas the observed ratio is about 2--3. Given the uncertainty
of any estimates of the amount of depolarization, we cannot exclude
that models with a non-uniform distribution of cosmic rays could be reconciled with
observations.

\begin{figure}
\begin{center}
\includegraphics[width=0.48\textwidth]{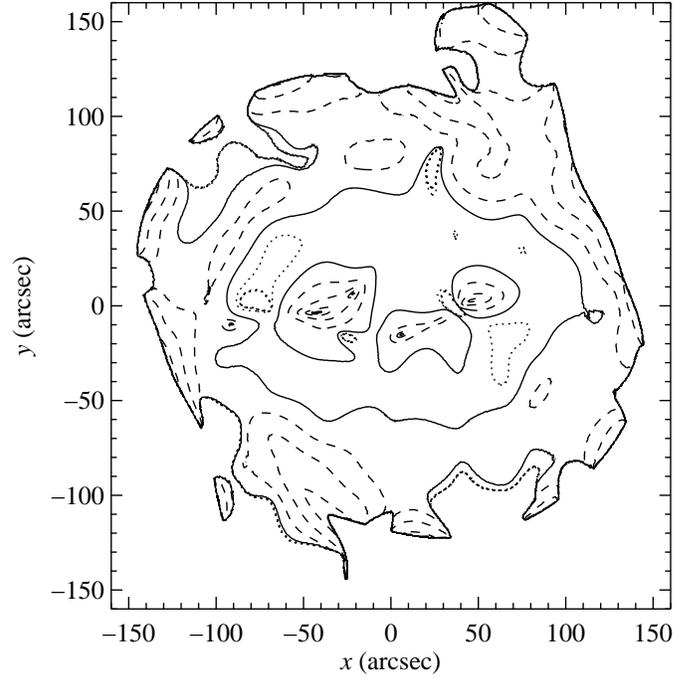}
\end{center}
\caption{\label{ncr}A map of the ratio of polarized intensity observed at
$\lambda6.2\cm$ to the synthetic $\PI$ obtained from Model 2 with
$\ncr=\mathrm{const}$. This ratio can be used to assess the variation of
$\ncr$ required to achieve a perfect fit to observations
(neglecting any anisotropy of the turbulent magnetic fields and
any contribution of a hypothetical galactic halo to the polarized emission).
The contour at level
unity is shown solid, the other contours are at levels $3^n$, with
negative and positive integer values of $n$; contours above (below) unity are
shown dashed (dotted).}
\end{figure}

We show in Fig.~\ref{ncr} the ratio of polarized
intensity observed at $\lambda6.2\cm$ to the integral along the line of sight
$\int B_\perp^2\,ds$, with $B_\perp$ the component of the modelled magnetic
field in the sky plane. The latter is proportional to the
synthetic polarized intensity obtained for a uniform cosmic ray distribution.
If our magnetic field model were perfect, the above ratio would show the
variation of cosmic rays across the galactic image. We note that the value of
the ratio varies remarkably little in the bar region; the most prominent
variations arise from the local peaks of the observed polarized intensity that
are also prominent in Fig.~\ref{diff_map}. Figure~\ref{ncr} confirms that the
variation of cosmic ray energy density within the galaxy is rather weak and
consists of a large-scale, smooth variation with contours of a shape similar
to that of gas density and other tracers in the bar, and perhaps with a few
local maxima.

\subsection{The effects of the turbulent magnetic diffusivity}
Models~2 and 4 yield very similar magnetic field
distributions (see Fig.~\ref{ThirtyDegCuts}),
even though the
background turbulent magnetic diffusivity $\eta_0$ in Model 4 is more
than twice as large as in Model 2 (see Table~\ref{run_params}).
The main effect of enhanced magnetic diffusivity is to make
the secondary peaks of $\PI$ in Model~4 less prominent, even with
$\ncr=\mbox{const}$. Model~4 could be reconciled with observations
if $\ncr$
were enhanced in the outer bar regions and in the spiral arms and/or reduced in
the central part.
Unless this is the case, and given that Model~2 agrees with observations
better than Model~4, we conclude that our models support a value
$\eta_0\la10^{26}\cm^2\s^{-1}$ in NGC~1365.

\begin{figure}
\begin{center}
\includegraphics[width=0.4\textwidth]{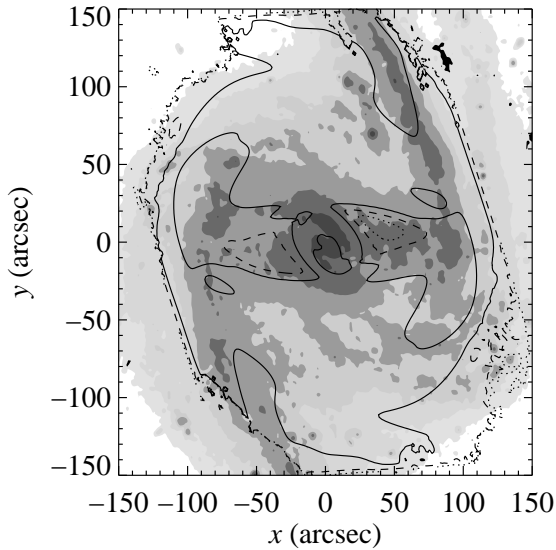}
\end{center}
\caption{\label{alignment}The alignment of the regular magnetic and
velocity fields (in the non-rotating frame) illustrated with the
contours of $\cos\chi=|\vec{u}\cdot\vec{B}|/uB$ at the midplane for Model 2,
projected on to the optical image in the plane of the sky. The
levels shown are 0.4 (dotted), 0.7 (dashed) and 0.95 (solid).}
\end{figure}

One of the effects of turbulent magnetic diffusion (and dynamo action) is to
produce a misalignment between magnetic field and velocity in a shearing flow.
Since the velocity shear is strong everywhere in the bar region and near the
spiral arms, we expect that magnetic and velocity fields would be tightly
aligned (in the corotating frame) if magnetic field were frozen into the flow
(Moss et al.\ \cite{MS01}).

The degree of alignment between the model magnetic and velocity fields in
NGC~1365 is illustrated in Fig.~\ref{alignment}, where we note that the
angle between the two vectors exceeds $20^\circ$ almost everywhere in the bar.
The misalignment is also significant near the spiral arms. 

The local enhancements of turbulent magnetic diffusivity, by a factor of
2--3, in the dust lanes and near the galactic centre introduced in
Sect.~\ref{TDM} are important in our model as they allow us to avoid excessively
large strength of the large-scale magnetic field produced by extreme velocity
shear in those regions. As argued by Moss et al.\ (\cite{MS01}), such a local
enhancement of interstellar turbulence may be associated with instabilities of
the shear flow.

\subsection{The role of dynamo action}\label{role}
Model~3 has the same parameters as Model~2, except that $R_\alpha=0$. Thus,
this model includes the same effects of rotation and velocity shear as Model~2,
but without any dynamo action. In this model, the magnetic field decays on a
timescale
of about $0.6\Gyr$, after an
initial accelerated decay near the galactic centre.
However, the magnetic structure in the outer parts of the galaxy is remarkably
similar to that with $R_\alpha\neq0$. We deduce that the magnetic structure
we have obtained
does not depend strongly on details of
the poorly known $\alpha$-effect, but rather is
controlled by the large-scale velocity field, which is known much more
reliably. The role of the $\alpha$-effect is just to maintain
the magnetic field
against decay, which is enhanced by the strong shear typical of barred galaxies.
This situation is similar to that found when
modelling another barred galaxy, NGC 1097
(Moss et al.\ \cite{MS01}).

\subsection{The effect of the speed of sound}\label{TEOTSS}
Models~2 and 6 have different values for the speed of sound (10 and
$30\kms$, respectively). The higher speed of sound results in less structure in
the velocity and density fields (even though we neglected to include an
equivalent
increase of the turbulent magnetic diffusivity by a factor of 3, which would
be necessary for strict consistency). The cuts for Model~6 have
relatively weaker features in the
outer parts of the galaxy. This result appears less
acceptable, and we deduce
that $10\kms$ is a more favourable value for the
speed of sound of the gas phase, to which the regular magnetic field is
coupled.

\subsection{Dependence on the gas ionization fraction}\label{ion_frac}

For $X \ga 0.2$, the
synthetic cuts of $\PI$ show a
much greater disagreement with the observed cuts than when $X=0.1$.
For example, for the cuts in Fig.~\ref{NinetyDegCuts}, the effect of increasing $X$
(and therefore increasing depolarization) is to broaden
the minima in $\PI$ about $\pm 50 \arcsec$, and increase the ratios of
the central maximum to outer maxima (at about $\pm 100 \arcsec$).

When we calculate synthetic $\RM$ for values of $X \la 0.01$, the range
of values obtained does not match the observed range
$\pm 600 \radm$. For example, at $X = 0.01$, the maximum synthetic
$\RM$ is about $350 \radm$.

Given that a constant ionization fraction is appropriate, our models
suggest that $0.01 \la X \la 0.2$ to be a plausible range of values.

\section{Discussion and conclusions}\label{Disc}

We have constructed a  three dimensional dynamo model
for NGC 1365, with the rotation curve, non-circular velocities
and gas density taken from a  dynamical model for this
particular galaxy. Thus, although we have taken a similar approach
as in our earlier
studies
of other specific barred galaxies (IC 4214, Moss, Rautiainen \& Salo \cite{MRS99};
NGC 1097, Moss et al.\ \cite{MS01}),
for NGC 1097 we adopted a generic dynamical model as input, whereas
here we have been able to use a bespoke model.
We have tried to make a much more detailed comparison between
observations and model predictions than previously -- but see also a recent paper 
by Vollmer et al. (\cite{V06}).

Of course, we have been restricted to using a mean field dynamo model --
for this sort of study there is really no plausible alternative
currently available.
Our modelling (and that of the earlier papers, cited above)
has demonstrated that when modelling galaxies with strong non-circular velocities
the role of the mean field $\alpha$-coefficient is primarily to offset the
inevitable diffusive decay of the field, and thus allow a steady
state with fields of order equipartition strength to be maintained.
The major determinant of the field structure is the non-circular
velocity field (Sect.~\ref{role} and Moss et al.\ \cite{MK98,MS01}), and the
main features can be expected to persist for plausible
field maintenance mechanisms.
We must further bear in mind the other limitations of the modelling,
including the restriction to the inner part of the galaxy, $r\le 15\kpc$,
which means that boundary effects may influence results near this radius.

Our main conclusions are as follows.
We see no evidence for a variation in $\ncr$ as strong as in $B^2$.
This may imply that equipartition between cosmic rays and the regular magnetic
field is {\it not\/} maintained even at global scales. The discrepancy
between our crude models involving the equipartition assumption and observations
can be significantly reduced if Faraday depolarization due to turbulent magnetic fields
is taken into account.  We have discussed this result further in Sect.~\ref{n_cr_models}.

The strongest deviations of the synthetic polarized intensity from that
observed occur in the bar region, just upstream of the dust lanes. The reason
for the low synthetic polarized intensity is the small value of  magnetic
field strength there in the model. More precisely,
the depth of the minima in synthetic and observed $\PI$ are similar
but they occur at somewhat different positions,
and those in the synthetic map are broader (see Fig.~\ref{NinetyDegCuts}).
However, the relative heights of the maxima in $\PI$
are reproduced quite successfully.
In this sense, the agreement is better than might be
inferred from Fig.~\ref{diff_map}. The reason for the difference is
the deep and broad minimum in the gas density in those regions.
We believe this to be a shortcoming of the gas dynamical model,
which was fitted to incomplete CO data. In particular,
CO observations of NGC~1097 (Crosthwaite \cite{CROS01}) do not
show the minima of the density in
the bar region to be as deep as in
the model of NGC~1365 used here.

Rather surprisingly, the agreement between the {\it orientations} of the
model magnetic field and the $B$-vectors derived from the observed polarization 
vectors is not as good as that between the model and synthetic PI distributions.
We have discussed this in Sect.~\ref{orientation}. The reason for this
difference remains unclear.

Our preferred model relies on the galactic rotation curve and gas density
distribution different from those suggested by Lindblad et al.\ \cite{LLA96};
the rotation curve used is that resulting from CO observations (Sofue at al.\
\cite{S99}). Our results are compatible with the observed distributions of
polarized synchrotron intensity and the magnitude of the Faraday rotation measure
for the number densities of ionized diffuse gas of order $0.16\cm^{-3}$ at
a distance of order $5\kpc$ from the centre along the bar's minor axis and
$0.21\cm^{-3}$ in the spiral arms. With the gas dynamical model used here,
this corresponds to the mean ionization fraction of 0.01--0.2.

Our models confirm that magnetic field strengths in the inner bar
region can be strong enough to drive mass inflow at a rate of several solar
masses a year (see also Moss, Shukurov \& Sokoloff \cite{MSS00}, Moss et al.\ \cite{MS01},
Beck et al.\ \cite{Betal05}). 
Thus, in these strongly barred, strongly magnetic
galaxies, it becomes necessary to include the dynamical effects of magnetic
fields in order to reproduce all features of the gas flow. It follows that
self-consistent magnetohydrodynamic modelling of barred galaxies is required.

Keeping in mind our dynamical model is incomplete, at least in that azimuthal
structure corresponding to odd modes is omitted,
our general conclusion is that mean field dynamo models are
reasonably
successful in modelling magnetic fields in this barred galaxy.
Moreover, such models can also provide information
about both the gas dynamical
modelling process and conditions in the interstellar medium
(see also the models for the `normal' spiral galaxy M31 in Moss et al.\ \cite{MS98}).
A robust conclusion is that, contrary to widely held opinions, dynamical
effects of magnetic fields cannot be everywhere ignored in galaxy modelling.

\appendix

\section{Synthetic radio maps}
\label{synthetic_radio_maps}

Synthetic radio maps have been obtained
by computing the Stokes parameters $Q$ and $U$
from the magnetic field obtained from the dynamo simulations.
The magnetic field was rotated to
the same position in the sky as the galaxy NGC~1365,
and $Q$ and $U$ at
a given wavelength were
obtained by integration along the line of sight (where the $z$-direction here
points towards an observer at infinity):
\[
Q= C {\int { \epsilon({\vec r}) \cos[2\psi({\vec r})]}}\; {dz},
\]
\[
U = C {\int { \epsilon({\vec r}) \sin[2\psi({\vec r})]}}\; {dz},
\]
with allowance for Faraday rotation by the thermal ionized in the local polarization
angle $\psi$ (see, for example, Sokoloff et al. (\cite{S98}) and references therein
for more details). Here $C$ is a dimensional constant whose specific value
is inessential here, and $\epsilon$ is the synchrotron emissivity.
We assume that the synchrotron spectral index is a constant,
$q=-1$, so that
$\epsilon\propto\ncr B_\perp^2$. We use models of $\ncr$ from Sect.~\ref{n_cr_models}.

Depolarization due to differential Faraday rotation is included as we take
\[\label{local_polang}
\psi({\vec r}) = \psi_{0}({\vec r}) + C_1 \lambda^{2} {\int_{z}^{\infty}{\nel({\vec
r}) B_{\parallel}({\vec r})}\,dz},
\]
where $\lambda$ is the wavelength of the emission and
$C_1=0.81\rm{rad}\,\rm{m}^{-2}\cm^3\mkG^{-1}\p^{-1}$ is a
dimensional constant. The number density of
thermal electrons $\nel$ was obtained from the total gas
number density $n$
as given by the gas dynamical model, assuming a constant ionization degree of
$X$.
(So we take $\nel = X \rho_\mathrm{gas}/m_\mathrm{H}$ with $m_\mathrm{H}$ the
proton mass.)

The synthetic Stokes parameters are then given by convolving the map of raw
parameters with a Gaussian beam in sky plane.

\begin{acknowledgements}
We are grateful to E.~Athanassoula, P.~A.~B.~Lindblad and P.~O.~Lindblad for
providing their gas dynamical model of NGC~1365 and for useful discussions.
We thank A.~Fletcher for numerous useful discussions.
This work was supported by PPARC Grants PPA/G/S/2000/00528 and PPA/S/S/2000/02975A. 
DM, AS and DDS acknowledge the hospitality and financial support of the Isaac Newton Institute
for Mathematical Sciences (University of Cambridge) in 2004.
RB and DDS acknowledge financial support from DFG-RFBR project 03-02-04031.
DDS is grateful to RFBR for financial support under grant 04-02-16094.
This work was partially supported by Swiss Nationalfonds grant 200020-101766.
\end{acknowledgements}



\end{document}